\documentclass{pas}

\usepackage{multirow}

\usepackage{xcolor}
\usepackage{comment}

\newcommand{\Oiiiu}{O~{\sc iii}]~\ensuremath{\lambda 1666}}
\newcommand{\Oiiiuo}{O~{\sc iii}]~\ensuremath{\lambda 1661}}
\newcommand{\Oiiit}{[O~{\sc iii}]~\ensuremath{\lambda 4363}}
\newcommand{\Oiiin}{[O~{\sc iii}]~\ensuremath{\lambda 4959}}
\newcommand{\Oiiis}{[O~{\sc iii}]~\ensuremath{\lambda 5007}}
\newcommand{\Heiiu}{He~{\sc ii}~\ensuremath{\lambda 1640}}
\newcommand{\Heiio}{He~{\sc ii}~\ensuremath{\lambda 4686}}

\newcommand{\Ciiiu}{C~{\sc iii]}~\ensuremath{\lambda\lambda 1907, 1909}}
\newcommand{\Civ}{C~{\sc iv}~\ensuremath{\lambda\lambda 1549, 1551}}

\newcommand{\Siios}{[S~{\sc ii}]~\ensuremath{\lambda 6716}}
\newcommand{\Oiiiuvratio}{{O~{{\sc{iii}}}]~\ensuremath{\lambda 1666}}~/~{[O~{\sc{iii}}]~\ensuremath{\lambda 4959}}}
\newcommand{\Oiiioptratio}{{[O~{{\sc{iii}}}]~\ensuremath{\lambda\lambda 4363}}~/~4959}
\newcommand{\Teuv}{$T_{e~1666}$}
\newcommand{\Teopt}{$T_{e~4363}$}
\newcommand{\densratio}{[S~{\sc ii}]~\ensuremath{\lambda\lambda}6731~/~6716}
\newcommand{\Opp}{O$^{++}$}

\newcommand{\Hp}{H$^{+}$}
\newcommand{\Hepp}{He$^{++}$}

\newcommand{\Hi}{H~{\sc i}}
\newcommand{\Hii}{H~{\sc ii}}
\newcommand{\Hei}{He~{\sc i}}
\newcommand{\Heii}{He~{\sc ii}}

\newcommand{\Oii}{[O~{\sc ii}]}
\newcommand{\Oiii}{[O~{\sc iii}]}
\newcommand{\Oiiib}{O~{\sc iii]}}
\newcommand{\Neiii}{[Ne~{\sc iii}]}

\newcommand{\Ha}{H$\alpha$}
\newcommand{\Hb}{H$\beta$}
\newcommand{\Hg}{H$\gamma$}
\newcommand{\Hd}{H$\delta$}

\newcommand{\Sii}{[S~{\sc ii}]}

\newcommand{\kms}{km~s$^{-1}$}

\newcommand{\farcs}{\ensuremath{^{\prime\prime}}}
\newcommand{\arcsec}{\ensuremath{^{\prime\prime}}}

\defcitealias{Senchyna2017}{S17}
\defcitealias{Senchyna2022}{S22}
\defcitealias{Mingozzi_etal_2022}{M22}

\begin{document}

\lefttitle{UV and Optical \Hii\ Region Temperatures}
\righttitle{E. Huntzinger \textit{et al.}}

\jnlPage{1}{4}
\jnlDoiYr{2025}
\doival{10.1017/pasa.xxxx.xx}

\articletitt{Research Paper}

\title{Consistent Gas-Phase Temperatures and Metallicities from UV and Optical Nebular Emission: A Reliable Foundation from z=0 to Cosmic Dawn}

\author{\gn{Erin} \sn{Huntzinger}$^{1}$, \gn{Yuguang} \sn{Chen}$^{2,1}$, \gn{Tucker} \sn{Jones}$^{1}$, \gn{Ryan} \sn{Sanders}$^{3}$, \gn{Peter} \sn{Senchyna}$^{4}$, \gn{Daniel P.} \sn{Stark}$^{5}$, \gn{Fabio} \sn{Bresolin}$^{6}$, \gn{Stephane} \sn{Charlot}$^{7}$ and \gn{Jacopo} \sn{Chevallard}$^{8}$}

\affil{$^{1}$ Department of Physics and Astronomy, University of California Davis, 1 Shields Avenue, Davis, CA 95616, USA\\
$^{2}$ Department of Physics, The Chinese University of Hong Kong, N.T., Hong Kong SAR, China\\
$^{3}$ Department of Physics and Astronomy, University of Kentucky, Lexington, KY 40506, USA\\
$^{4}$ The Observatories of the Carnegie Institution for Science, 813 Santa Barbara Street, Pasadena, CA 91101, USA\\
$^{5}$ Department of Astronomy, University of California, 501 Campbell Hall \#3411, Berkeley, CA 94720, USA \\
$^{6}$ Institute for Astronomy, University of Hawaii, Honolulu, HI 96822, USA\\
$^{7}$ Sorbonne Universit\'e, CNRS, UMR 7095, Institut d’Astrophysique de Paris,
98 bis bd Arago, F-75014 Paris, France\\
$^{8}$ Department of Physics, University of Oxford, Denys Wilkinson Building, Keble Road, Oxford OX1 3RH, UK\\}

\corresp{Yuguang Chen, Email: yuguangchen@cuhk.edu.hk}

\citeauth{Huntzinger E., Chen Y., Jones T., Sanders R., Senchyna P., Stark D.P., Bresolin F., Charlot S. and Chevallard J. (2025) Consistent Gas-Phase Temperatures and Metallicities from UV and Optical Nebular Emission for \Heii\ Emitting Galaxies. {\it Publications of the Astronomical Society of Australia} {\bf 00}, 1--12. https://doi.org/10.1017/pasa.xxxx.xx}

\history{(Received xx xx xxxx; revised xx xx xxxx; accepted xx xx xxxx)}

\begin{abstract}
The rest-frame ultraviolet (UV) spectra of star-forming galaxies are increasingly important as they become one of the primary windows to probe the physical properties of cosmic dawn ($z>8$) galaxies with the James Webb Space Telescope. However, the systematic discrepancies between UV and optical gas-phase metallicity measurements remain poorly understood in the local universe, partly due to challenges in achieving precise comparisons between UV and optical spectra for the same objects. In this work, we introduce a novel method that leverages the \Heiiu\ and \Heiio\ nebular emission lines to achieve accurate aperture and reddening corrections between UV and optical spectra. Here we apply this method to three nearby Blue Compact Dwarf (BCD) galaxies. Our results demonstrate that this approach enables precise measurements, with electron temperatures ($T_e$) derived from UV and optical spectra exhibiting closer agreement compared to previous studies, and O/H abundance agreeing within 0.1 dex. However, two BCDs appear to have lower UV-based electron temperatures \Teuv~$<$~\Teopt, in contrast to expectations from {the temperature fluctuation model}.
We consider a variety of possible explanations for these unphysical temperatures -- differential dust attenuation, aperture differences, and spatial extent of emission lines -- but no suitable cause is identified.
These findings suggest a complex gaseous environment associated with star formation, and underscore the need for additional observations to further investigate the nature of \Heii\ nebular emission and address the systematic issues between UV and optical nebular properties.
Nonetheless, the close empirical agreement of these results indicates that UV- and optical-based nebular {temperature and abundance} measurements can be reliably compared within 0.1 dex, providing a solid foundation for evolutionary studies from the local Universe to cosmic dawn.
\end{abstract}


\maketitle

\section{Introduction}

The processes of gas accretion, star formation and gas outflows are paramount in understanding how galaxies form and evolve. Accurate measurements of gas-phase chemical composition provide some of the most robust constraints on these processes \citep[e.g.,][]{Erb2006,Erb2008,Peeples&Shankar2011,Lilly2013,Dave2011,Dave2017,Sanders2021}. Gas accretion dilutes metals in the interstellar medium (ISM), star formation produces metals through nucleosynthesis and then returns them to the ISM through supernovae and stellar winds, and outflows remove metals from galaxies. 
Thus, the gas-phase metallicity of a galaxy is regulated by the interplay between gas inflows, outflows, and star formation.
Consequently, studies of chemical evolution serve as a powerful tool to infer the baryonic processes shaping galaxies over cosmic time.

Gas-phase oxygen abundance (referred to herein as `metallicity,' expressed as $12+\log{(\mathrm{O/H})}$) has now been measured in galaxy populations reaching $z>8$ with the James Webb Space Telescope \citep[JWST; e.g.,][]{ArellanoCordova2022,Schaerer2022,Taylor2022,Bunker2023,Curti2023,Nakajima2023,Sanders2023,Trump2023,Sarkar2025}. 
These advancements build upon extensive earlier surveys that have mapped metallicity evolution across $z \simeq 0$--4 \citep[e.g.,][]{Kriek2015,Steidel2014,Wisnioski2015,Stott2016,Treu2015,Momcheva2016}. 
These efforts have consistently found lower metallicities at higher redshifts and at lower stellar masses \citep[e.g.,][]{Tacconi2018,Sanders2021,Tremonti2004,Cullen2014,Curti2020,Troncoso2014,Andrews&Martini2013,Erb2006,Maiolino2008}. 
The observed mass-metallicity relation and its evolution are primarily attributed to a mass-dependent outflow rate, and higher gas fractions at earlier cosmic epochs.

The most common approach to measuring the O/H {abundances} relies on nebular emission lines originating in {\Hii} regions.
Collisionally excited lines (CELs) from \Oiii\ are particularly sensitive to the electron temperature ($T_e$). 
The $T_e$ can be determined from the ratio of two CELs probing different {upper energy levels} (e.g., \Oiiit/\Oiiis), and used to calculate the emissivity of oxygen and {other} emission lines, allowing for a precise determination of O/H abundance.
This technique, known as the `direct $T_e$' method, is commonly applied to rest-frame optical observations. However, recent JWST observations of galaxies at $z \gtrsim 8$ increasingly rely on rest-UV nebular emission lines (e.g., \Oiiib~$\lambda\lambda$1661,1666, \Ciiiu, \Civ) to probe gas-phase properties \citep[e.g.,][]{curti2025,Hayes2025,Hsiao2025,Tang2025,ArellanoCordova2026}, as the optical lines are redshifted into the more observationally challenging mid-infrared, particularly observations at $z \gtrsim 10$. 
Compared to the extensive optical surveys, rest-UV spectroscopic datasets at low redshift remain more limited due to atmospheric opacity below $\sim 3000\,\mathrm{\AA}$ and the necessary reliance on space-based facilities with limited sensitivity. Recent programs with the Hubble Space Telescope such as the COS Legacy Archive Spectroscopic SurveY (CLASSY; \citealt{Berg2022, James2022}) have substantially improved the availability of high-quality UV spectra for nearby star-forming galaxies, enabling the development of UV-based nebular diagnostics and providing insights into the ionizing stellar populations that shape UV line ratios. However, precise and widely calibrated $T_e$ measurements based on UV collisionally excited lines remain comparatively scarce relative to their optical counterparts.

Another challenge for chemical evolution studies is that even in sensitive rest-optical spectra, two `direct' methods which can be used to derive the {\Opp}/{\Hp} nebular abundance in an {\Hii} region systematically disagree. O/H abundances from CELs using the direct-$T_e$ method are found to be systematically lower (by $\sim$0.2-0.3 dex) compared to measurements from the much fainter {oxygen} recombination lines (RLs). 
This {so-called} Abundance Discrepancy Factor (ADF; e.g., \citealt{Tsamis2003, Blanc2015, Esteban2009, Esteban2014,GarciaRojas2004,Garcia2007}) is often attributed to temperature fluctuations of the gas within {\Hii} regions \citep{Peimbert67,GarciaRojas2004}.
In this scenario, $T_e$ values measured from optical CELs are biased high resulting in the direct-$T_e$ metallicities biased low. Meanwhile, the recombination line metallicities are accurate due to their insensitivity to $T_e$. For the UV emission lines, such as {\Oiiib~$\lambda\lambda$1661,1666}, this would suggest that $T_e$ derived from rest-UV lines would be biased even higher and thus metallicities even lower compared to the optical CELs.
On the other hand, another explanation of the ADF involves inclusions of high-metallicity gas \citep{Croxall2013, Stasinska2007} {among an ambient medium with lower metallicity}, in which case the CELs are expected to be more accurate {than RLs}. Some studies have found good agreement between the metallicities of young stars and CEL-based nebular measurements \citep[e.g.,][]{bresolin2016,Bresolin2025}, suggesting that $T_e$ metallicities may be more {reliable}.
As the cause of the ADF is currently not conclusively established, empirical relationships are important to establish the relative biases between different metallicity measurement techniques, {such as the joint use of UV and optical CELs}.

Significant efforts have been made in recent years to understand the gas properties measured from UV nebular emission in the local universe \citep[e.g.,][]{kunth97, Senchyna2017, senchyna2020, Senchyna2022, Berg2022, Kelly2025}. However, measuring the $T_e$ and metallicity using {UV} emission lines is challenging, since it requires an accurate comparison between the UV and optical fluxes (e.g., \Ciiiu\ and \Hi\ Balmer lines {or using \Oiiiu~/~\Oiiis\ to derive $T_e$.}). Such comparisons often introduce systematic uncertainty from different instruments and apertures used for the UV and optical measurements \citep[e.g.,][]{ArellanoCordova2022}, in addition to highly uncertain reddening corrections.
Using the CLASSY survey, \citet{Mingozzi_etal_2022} explored the promise and limitations of combining Hubble Space Telescope (HST)/Cosmic Origins Spectrograph (COS) UV spectra and various optical spectra by matching the continuum spectra to the stellar population synthesis models. Their results suggest that the \Oiii\ $T_e$ measured from the UV emission {(via \Oiiib~$\lambda\lambda$1661,1666 / \Oiiis)} is consistent with the optical $T_e$ {(via \Oiiit\ / \Oiiis)} {on average} with a sample standard deviation scatter of $\sim$1500~K ($\sim$10\% of $T_e$),
in relatively good agreement, but with substantial uncertainties related to dust attenuation and other effects.

In this work, we {demonstrate} a novel method to achieve precise comparison between the optical and UV fluxes. A key goal is to minimize uncertainty associated with combining UV and optical features, including aperture corrections and dust attenuation. We achieve this using nebular \Heii\ emission lines, namely UV \Heiiu\ and optical \Heiio. 
While the ionization source of these emission lines in \Hii\ regions is still under debate \citep{Kehrig2011, Kehrig2015, Schaerer2019,senchyna2020}, the production mechanism of nebular \Heii\ emission is believed to be dominated by recombination processes \citep{Schaerer98} analogous to the \Hi\ Balmer and Paschen lines. 
Accordingly, the flux ratio \Heiiu/\Heiio\ $\approx 6.99$ is expected to be nearly constant and relatively insensitive to gas density and $T_e$. This provides a valuable reference to calibrate observed fluxes from different instruments at rest-frame UV and optical wavelengths.
{In this work, we demonstrate that this method can provide precise measurements of $T_e$ and metallicity, and examine whether gas properties measured from UV features are systematically different from equivalent optical diagnostics.}

This paper is organized as follows. We present target galaxy selection and observations in Section \ref{sec:data}. Section \ref{sec:analysis} describes our methodology and analysis, including flux calibration of optical and UV spectra, dust attenuation and aperture corrections, and measurements of the physical properties.
We discuss the reliability of our measurements and the physical implications in Section \ref{sec:discussion}. Section \ref{conclusion} summarizes our results. 
Emission line analyses (including correction for dust attenuation and determination of the physical properties of the gas) are performed using \textsc{Pyneb} \citep{Luridiana2015}. We adopt the collision strengths from \citet{Tayal2017} for the \Opp\ ion and the transition probabilities {for all relevant lines} from \citet{FroeseFischer_etal_2004} throughout this work.

\section{Sample Selection and Data}
\label{sec:data}

Here we describe the target sample and spectroscopic data used in this work.
Our sample was selected from a superset of ten metal-poor star-forming galaxies {analyzed} in \citet[][hereafter \citetalias{Senchyna2017}]{Senchyna2017}, which presented UV spectra of these galaxies from HST's Cosmic Origins Spectrograph (COS) along with optical spectra from Keck's Echelle Spectrograph and Imager \citep[ESI;][]{Sheinis2002}. These galaxies were initially selected based on the presence of prominent {\Heiio} emission in their optical SDSS spectra \citep{Shirazi&Brinchmann2012}. 
{Since our objective is to characterize UV and optical nebular emisson, we restrict our analysis to the sample of galaxies exhibiting clear {narrow} nebular {\Heiiu} and {\Heiio} emission (as opposed to broad stellar wind emission) in high-resolution HST/COS and Keck/ESI spectra. The combination of optical and UV {\Heii} lines is necessary to obtain precise dust reddening and aperture corrections (see Section~\ref{sec:UVreddening} for details).}

\begin{figure*}
\center
 \includegraphics[width=2.33in]{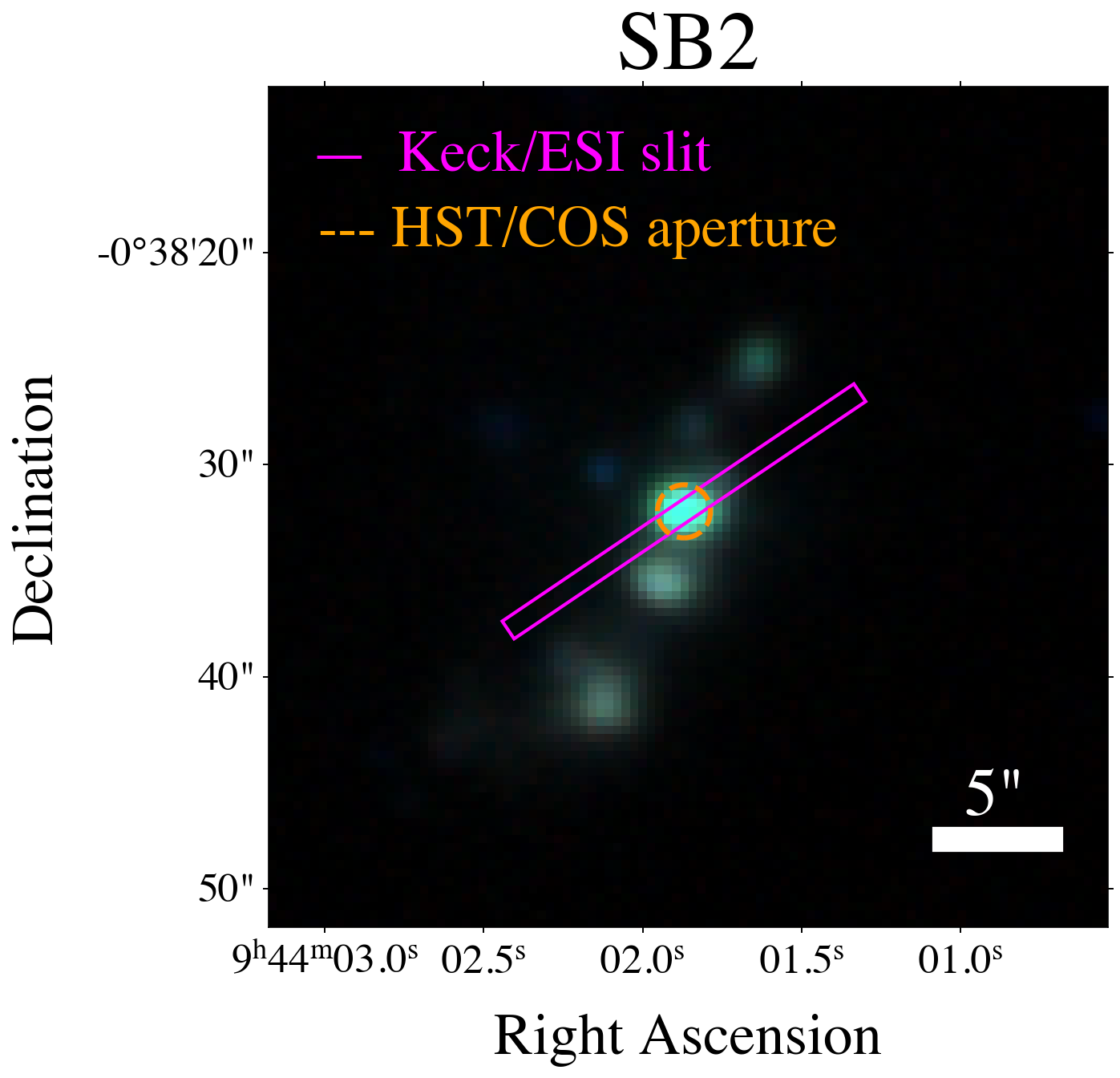}
 \includegraphics[width=2.295in]{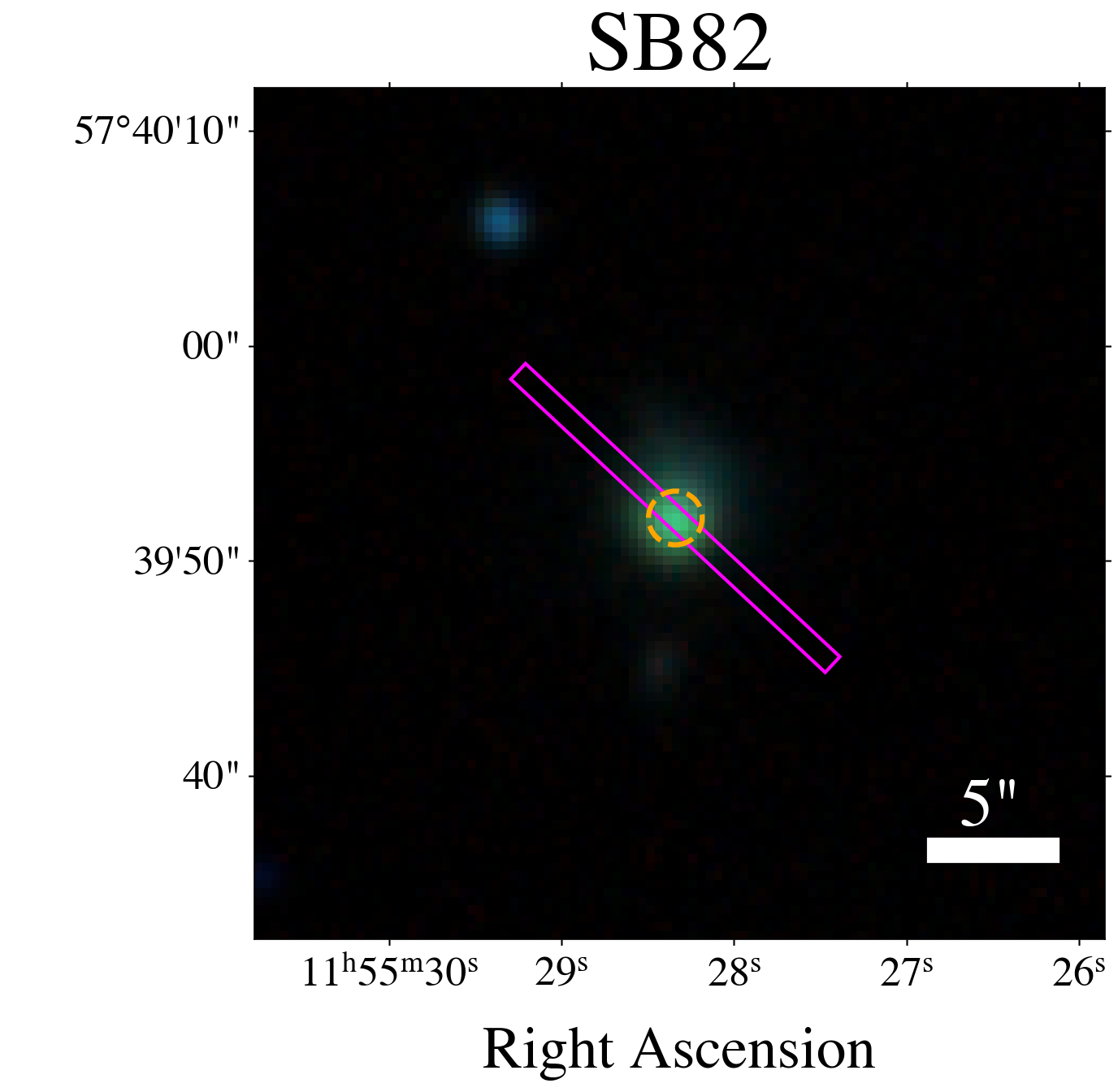}
 \includegraphics[width=2.3in]{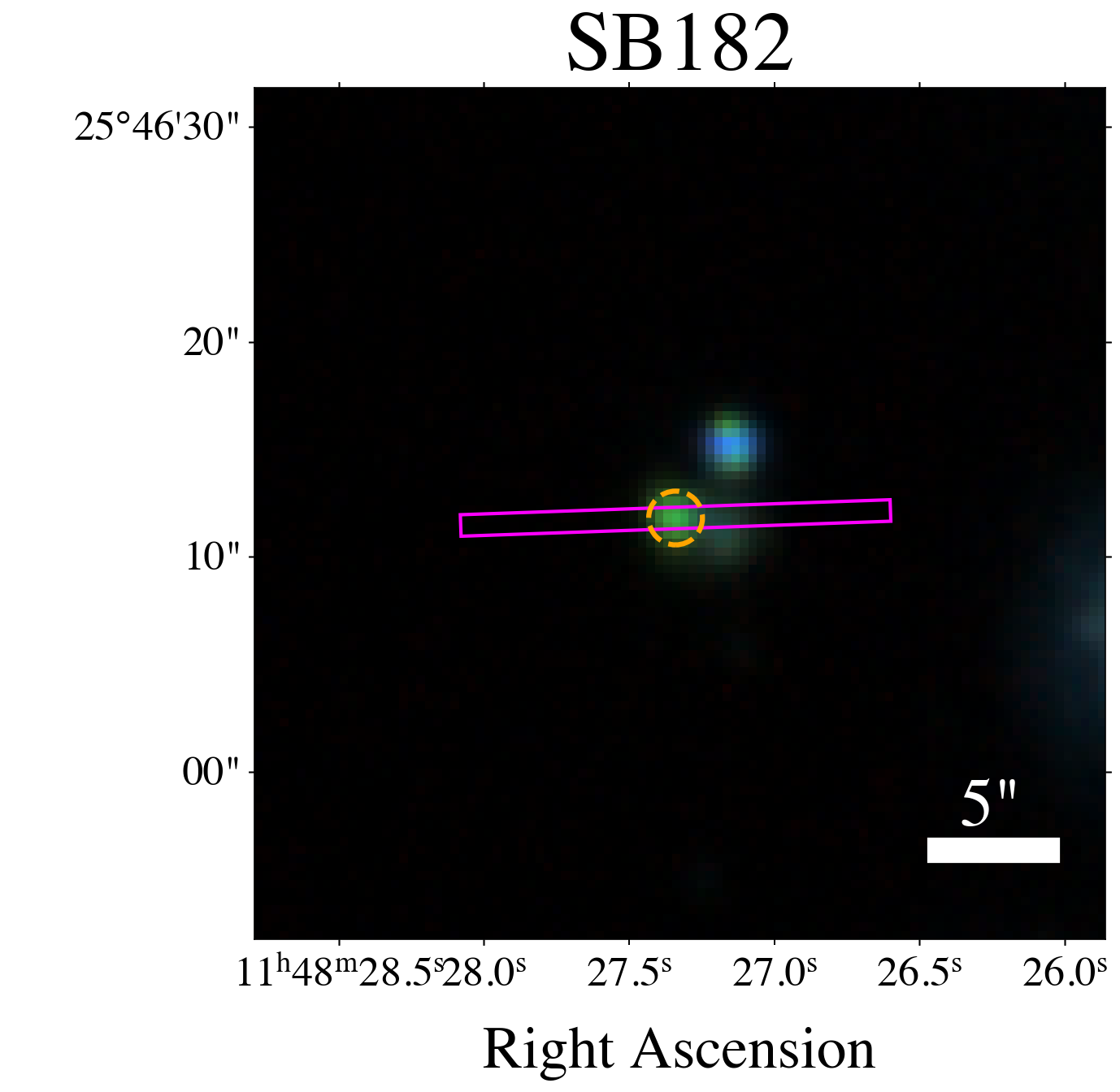}
 \caption{Images of the three blue compact dwarf galaxies studied in this work. Each panel shows a SDSS {\em{u, g, r}} false-color image, the {\em{HST/ COS}} 2\farcs5 aperture (circles), and the Keck/ ESI 1\arcsec\ $\times$ 20\arcsec\ slit (rectangles). 
 The 2\farcs5 {COS aperture size corresponds to projected diameters of} 0.23 kpc for SB~2, 0.92 kpc for SB~82, and 2.31 kpc for SB~182 \citep{Senchyna2017}. Near-UV target acquisition images from COS for all three objects can be found in \citetalias{Senchyna2017}.
 All three objects are dwarf galaxies containing one primary \Hii\ region, 
 from which the majority of nebular line emission is captured by both the COS and ESI apertures. }
  \label{fig:aperture}
\end{figure*}

The \Heii\ selection criteria result in a sample of three blue compact dwarf galaxies (BCDs) with relatively low metallicities (12 + log (O/H) $<$ 8.1) and recent star formation activity. The properties of these galaxies are summarized in Table~\ref{tab:properties}. Figure~\ref{fig:aperture} presents optical images of the three BCDs. Below, we briefly outline the physical characteristics of each target (masses from \citetalias{Senchyna2017}):
\begin{itemize}
\item {\bf{SB~2}} is a dwarf galaxy hosting multiple star-forming regions. Our analysis centers on the brightest star-forming region located in the central knot, as shown in Figure \ref{fig:aperture}, which has a stellar mass of $M_* = 10^{5.1}$ M$_{\odot}$. This region was captured in isolation within the COS aperture \citep[][hereafter \citetalias{Senchyna2022}]{Senchyna2022}. The values used in this work pertain exclusively to the region enclosed by the COS aperture and do not account for the contributions from neighboring star-forming regions in the galaxy. 
\item {\bf{SB~82}} is an isolated dwarf galaxy with a stellar mass of $M_*$=10$^{6.4}$ M$_{\odot}$, also known as Mrk~193. 
\item {\bf{SB~182}} is a dwarf galaxy with a stellar mass of $M_* = 10^{7.3}$ M$_{\odot}$. Morphologically, it consists of two distinct components separated by several arcseconds. In this study, we focus on the southeast star-forming region. The angular separation between the two components is sufficient to ensure that the northwest component was not included in the spectra analyzed here. 
\end{itemize}

\begin{table*}
	\centering
	\caption{Summary of target galaxy properties. \\
    }
	\label{tab:properties}
	\begin{tabular}{lcccr}
	\hline
        {\bf Object:}  &  {SB~2}$^e$ & {SB~82} & {SB~182}\\[4.pt]
        {Alternative ID:} &  {J0944-0038,~CGCG007-025} & {MRK~193} & {J1148+2546,~LEDA~36857}\\
        \hline
        RA & 9:44:01.87 & 11:55:28.34 & 11:48:27.34 \\
        DEC & -0:38:32.2 & 57:39:52.0 & 25:46:11.8 \\
        $z_{\mathrm{UV}}^a$ & 0.00486 & 0.01726 & 0.04515\\
        $\log (M_{*} / M_\odot$) $^c$ & 5.1 & 6.4 & 7.3\\
        $\log (\mathrm{SFR} / (M_\odot$~yr$^{-1}))$ {$^b$} & -1.49$\pm^{0.21}_{0.15}$ & - & +0.29$\pm^{0.18}_{0.15}$\\
        12+log(O/H) $^c$ & 7.81$\pm$0.07 & 7.91$\pm$0.04 & 8.01$\pm$0.04\\
        {r$_{50}$} {$^b$} & 0.984 & - & 0.874\\
        {$n_e$ (cm$^{-3}$)} {$^d$} & 158 & 189 & 184 \\
		\hline
	\end{tabular}
        \begin{footnotesize}
            \begin{flushleft}
                $^{\text{a}}$ Redshifts measured from \Oiiib~$\lambda\lambda$1661,1666 nebular emission lines.\\
                $^{\text{b}}$ Obtained from \cite{Berg2022}, where SFR was measured using \textsc{Beagle} SED fitting.\\
                $^{\text{c}}$ Obtained from \cite{Senchyna2017}. $M_*$ was measured from SED fitting with a typical uncertainty of 0.1 dex, and $12+\log\mathrm{(O/H)}$ was measured using the optical direct $T_e$ method. 
                \\
                $^{\text{d}}$ $n_e$ derived from the \densratio\ flux ratio for each object.\\
                $^{\text{e}}$ Values for SB~2 reflect the brightest star-forming region in the central knot of the galaxy, as opposed to the (larger) entire galaxy.
            \end{flushleft}
        \end{footnotesize}
\end{table*}

\subsection{Keck/ESI}
\label{sec:keck}

The ESI data were taken on 29 March 2016 (SB~82) and 20 January 2017 (SB~2 and SB~182). Sky conditions were clear with {0.8--1.2 arcsecond} seeing. 
The {1''} slit width was used, resulting in spectral resolution $R\sim4000$ (75~\kms\ FWHM) spanning 3900--10900~\AA. To minimize aperture loss due to atmospheric dispersion, all observations were taken with a slit position {near the parallactic angle} (PA=304 for SB~2, 227 for SB~82, and 272 for SB~182). 
The total integration time was 2 hours for SB~2 and 2.5 hours for SB~82 and SB~182, split into individual exposures of 30 minutes each. 
The spectra were reduced using the \textsc{ESIRedux} data reduction pipeline as described in \citetalias{Senchyna2017}.

\begin{figure*}
    \centering
    \includegraphics[width=6.25in]{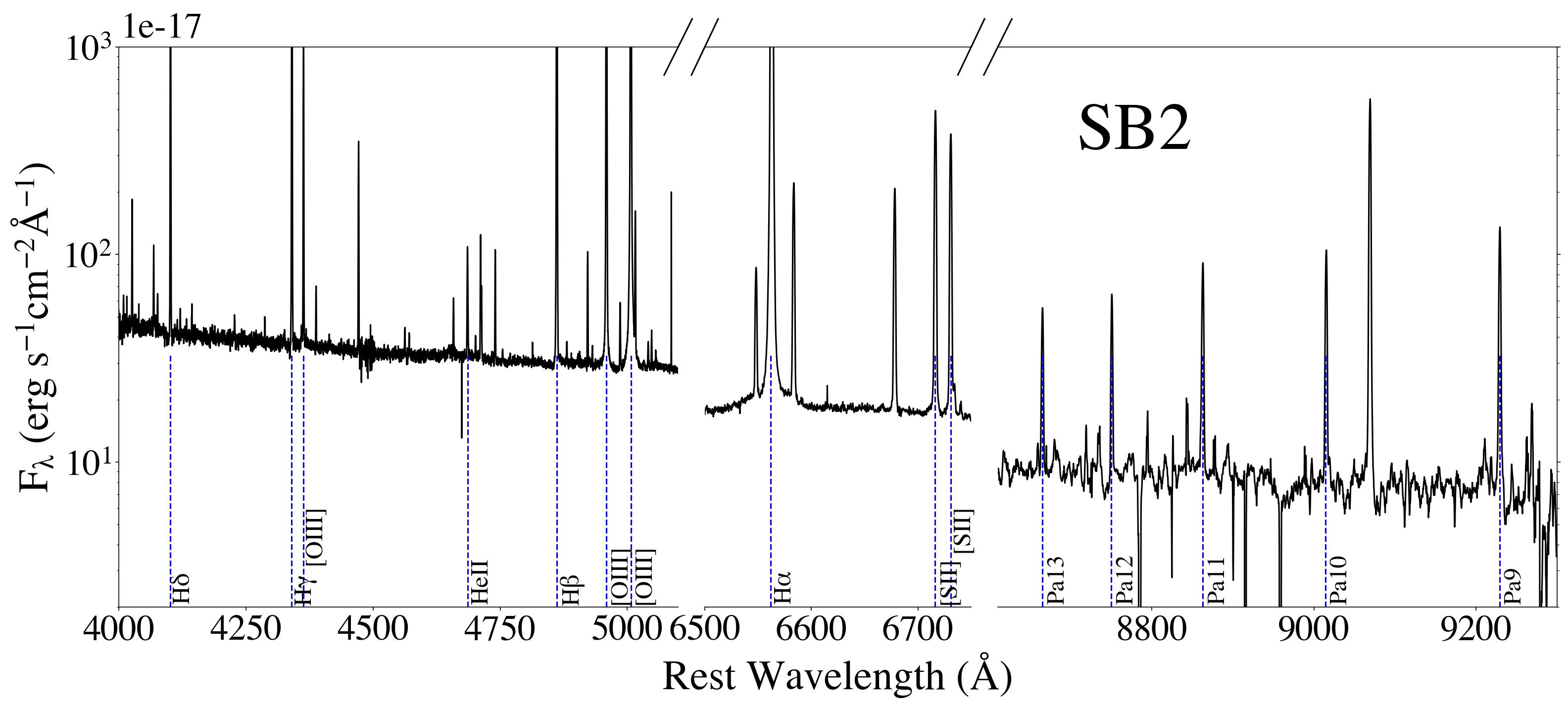}
    \vskip\baselineskip
    \includegraphics[width=6.25in]{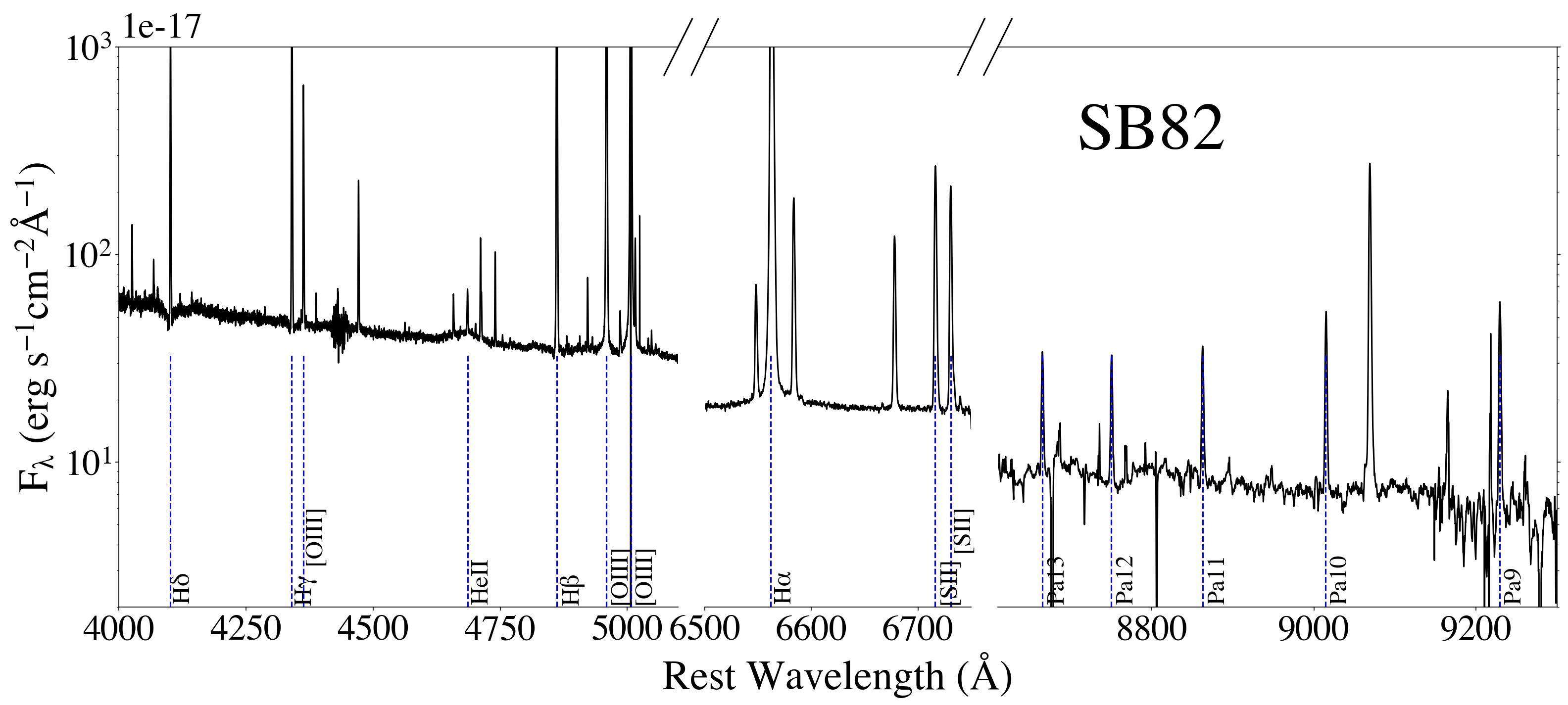}
    \vskip\baselineskip
    \includegraphics[width=6.25in]{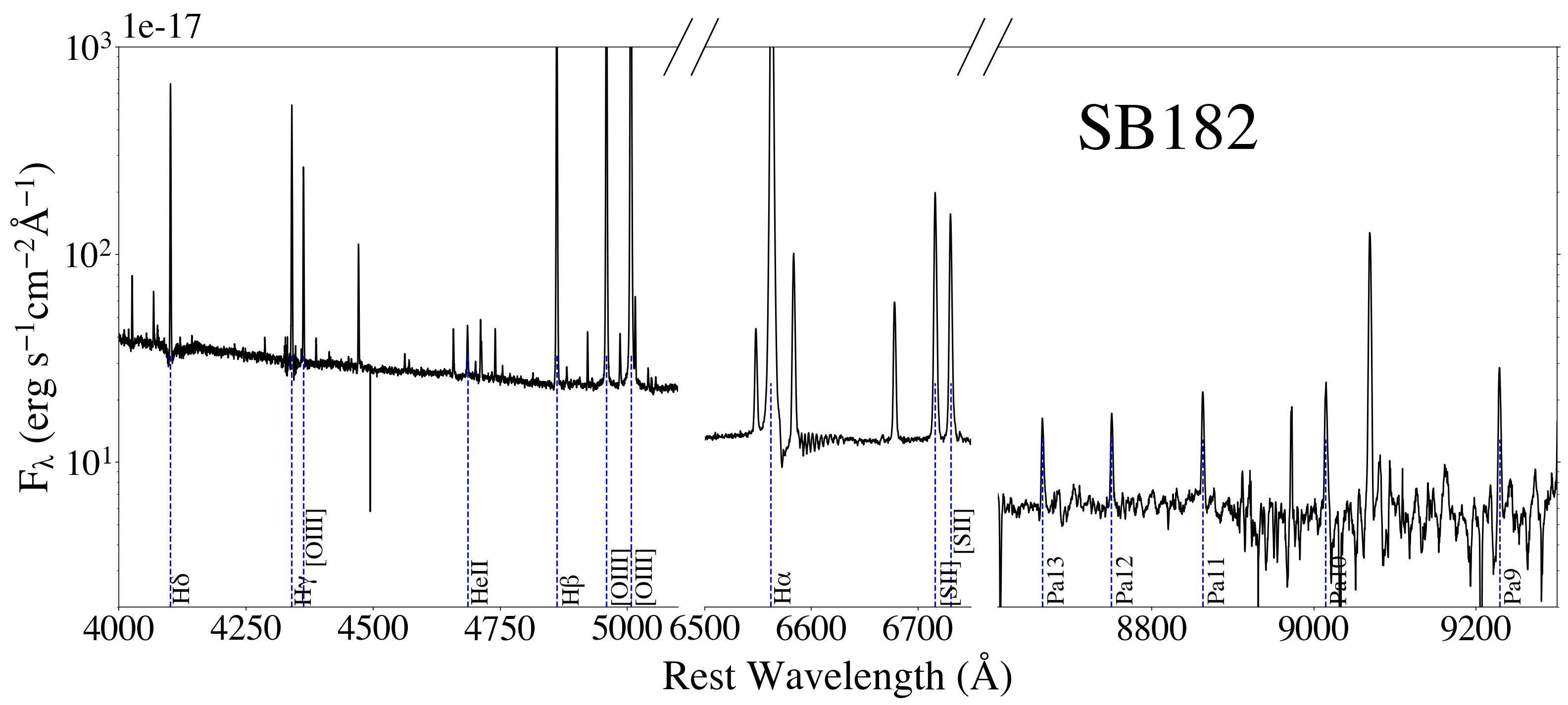}
    \caption{Keck/ESI optical spectra of the three objects presented in this work. Prominent emission features, epecially the ones useful for this work, are labeled {and marked with} vertical dashed lines. All relevant features are clearly detected, including nebular \Heiio. {A broad \Heii\ component is visible in SB~82.}}
    \label{fig:opt_flux_fits_full}
\end{figure*}

{To improve flux calibration relative to the \textsc{ESIRedux} processing,} we scale the continuum level of Keck/ESI spectra to match that in the fully calibrated spectra available for each target from the Sloan Digital Sky Survey \citep[SDSS;][]{York2000}.
We first masked out wavelength regions around the strong emission lines, and then  excluded pixels deviating more than $2\sigma$ from the running median in bins of 50~\AA\ to mask {detected spectral features}.
The masking of weak lines was necessary to robustly estimate the continuum due to the large number of emission lines detected ($>100$ per target) in these {sensitive} spectra.
We then fit a univariate cubic spline to the masked spectrum to obtain a smooth model of the continuum in the ESI spectrum.
We performed the same steps on the SDSS spectrum, {and divide the ESI by SDSS continuum model to obtain the flux calibration scaling factor. We then scale the ESI flux and error spectra by this factor.}
Figure \ref{fig:opt_flux_fits_full} shows the flux-calibrated ESI spectra of the three objects in the wavelength ranges most relevant for this work.

\subsection{HST/COS}
\label{sec:hst}

The HST/COS spectroscopic observations for the BCDs have been previously presented in \citetalias{Senchyna2017} and \citetalias{Senchyna2022}. Here we give a brief summary. The spectra were obtained using the 2\farcs5 COS aperture with the G160M grating centered at 1533 \AA.
For SB~2 and SB~82, the COS data were taken on 4--5 April 2020 (SB~2) and 21 November 2019 (SB~82) as part of the Cycle 26 HST-GO-15646 program. 
For SB~182, the data were taken on 05 November 2015 as part of the Cycle 23 HST-GO-14168 program. 
The total exposure times were 26,925.12 seconds for SB~2, 29,255.52 seconds for SB~82, and 2,624.61 seconds for SB~182.
The spectral resolution of the COS data is $\sim 15,000$.
Data for all three targets were reduced using the standard \textsc{Calcos} pipeline\footnote{\href{https://hst-docs.stsci.edu/cosdhb/chapter-3-cos-calibration/3-2-pipeline-processing-overview}{\textsc{Calcos} documentation}}, with the calibration files downloaded from the Space Telescope Science Data Analysis System (\textsc{STSDAS})\footnote{\href{https://www.stsci.edu/documents/dhb/pdf/c03_stsdas.pdf}{\textsc{STSDAS} documentation}}. For spectral extraction, we employed the ``TWOZONE'' algorithm. 
Additional details on data reduction and extraction for all three targets are provided in \citetalias{Senchyna2017} and \citetalias{Senchyna2022}.

\section{Analysis}
\label{sec:analysis}

The objectives of this work require computing the \Opp\ abundance from the optical \Oiiin~/~\Hb\ ratio using $T_e$ measured from the UV \Oiiiuvratio\ emission line ratio, and comparing it with $T_e$ derived from the optical \Oiiioptratio\ ratio.
This approach requires measuring accurate line fluxes from both the UV and optical spectra. Given the different excitation energies, \Oiiib~$\lambda\lambda$1661,1666 is more sensitive to $T_e$ than \Oiiit. Comparing these measurements can therefore provide constraints on temperature fluctuations within the ionized gas, which should cause the UV-based $T_e$ to {to be biased higher than the $T_e$ derived from optical \Oiiit}.
In this section we provide a detailed description of these measurements and methodology.

\subsection{Emission Line Fluxes}
\label{sec:em_line_fit}

\subsubsection{Optical Line Fitting}
\label{sec:opt_line_fit}

\begin{figure*}
    \centering
    \includegraphics[width=1.75in]{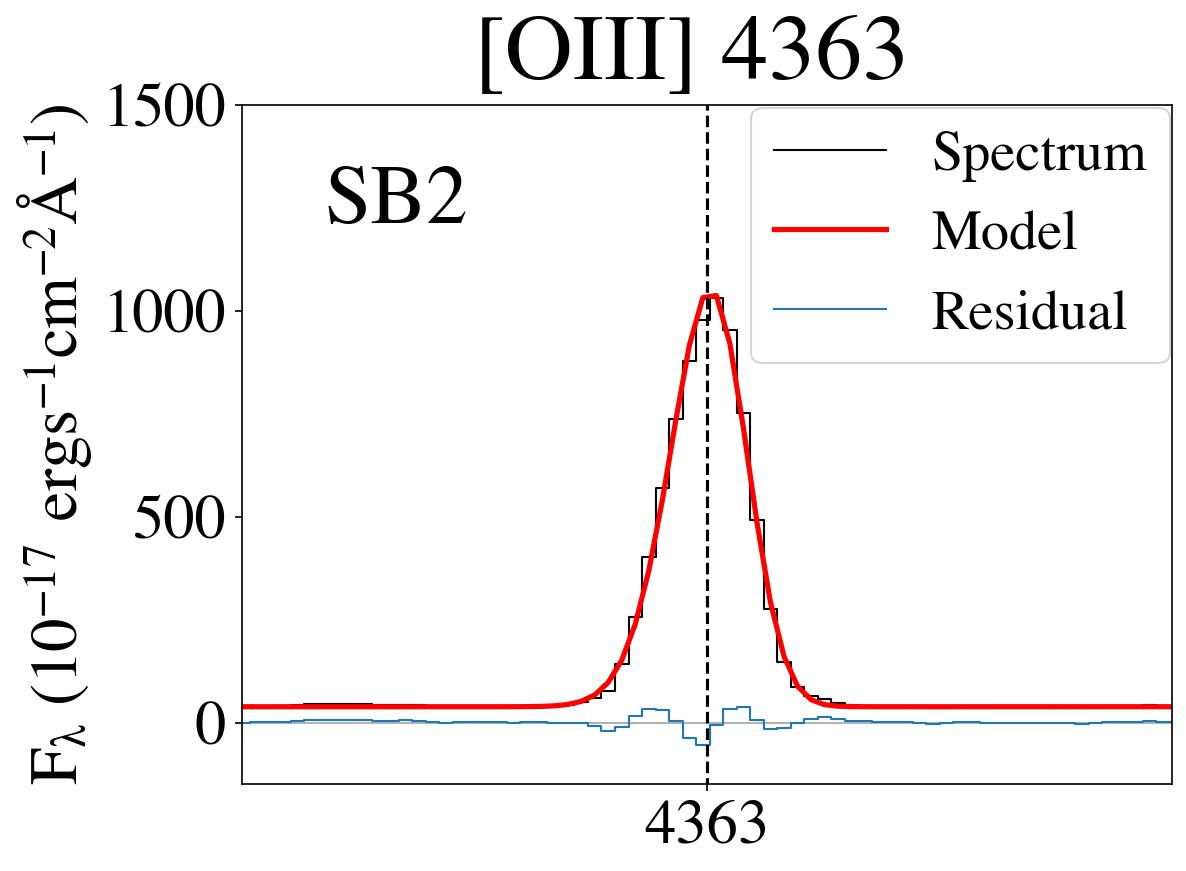} 
    \includegraphics[width=1.65in]{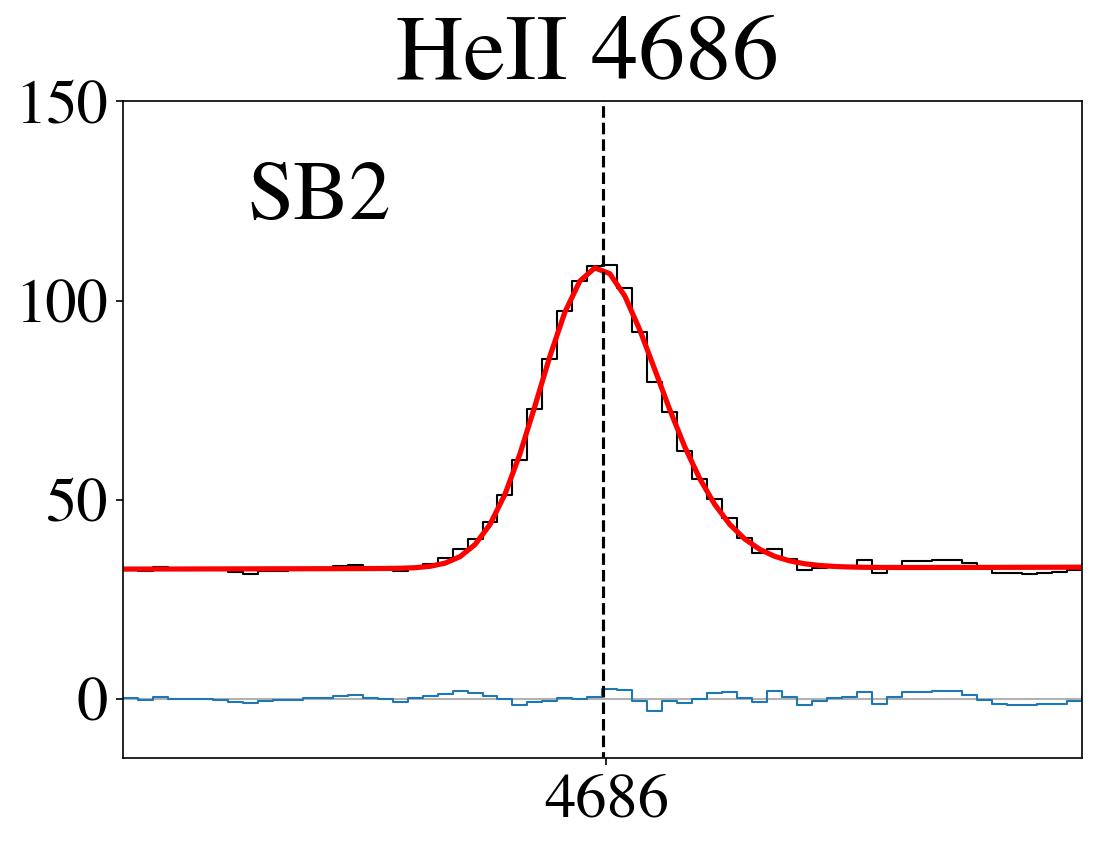}
    \includegraphics[width=1.65in]{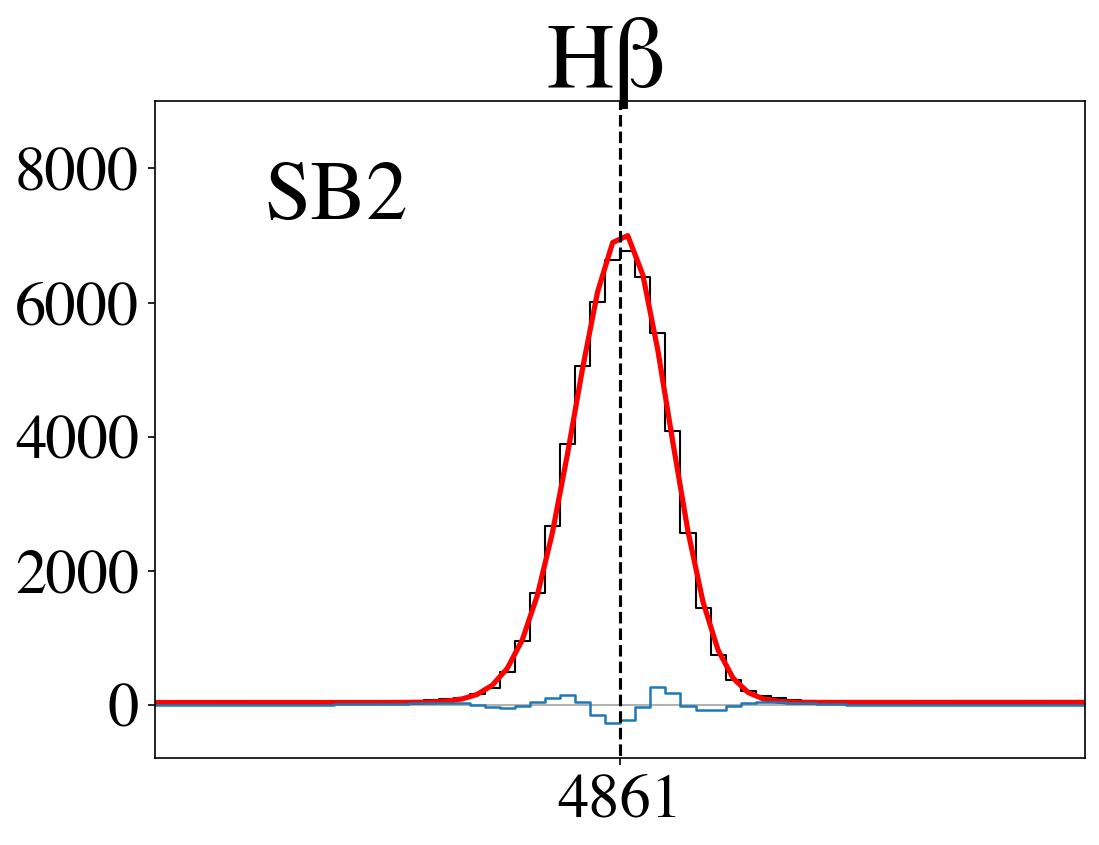}
    \includegraphics[width=1.65in]{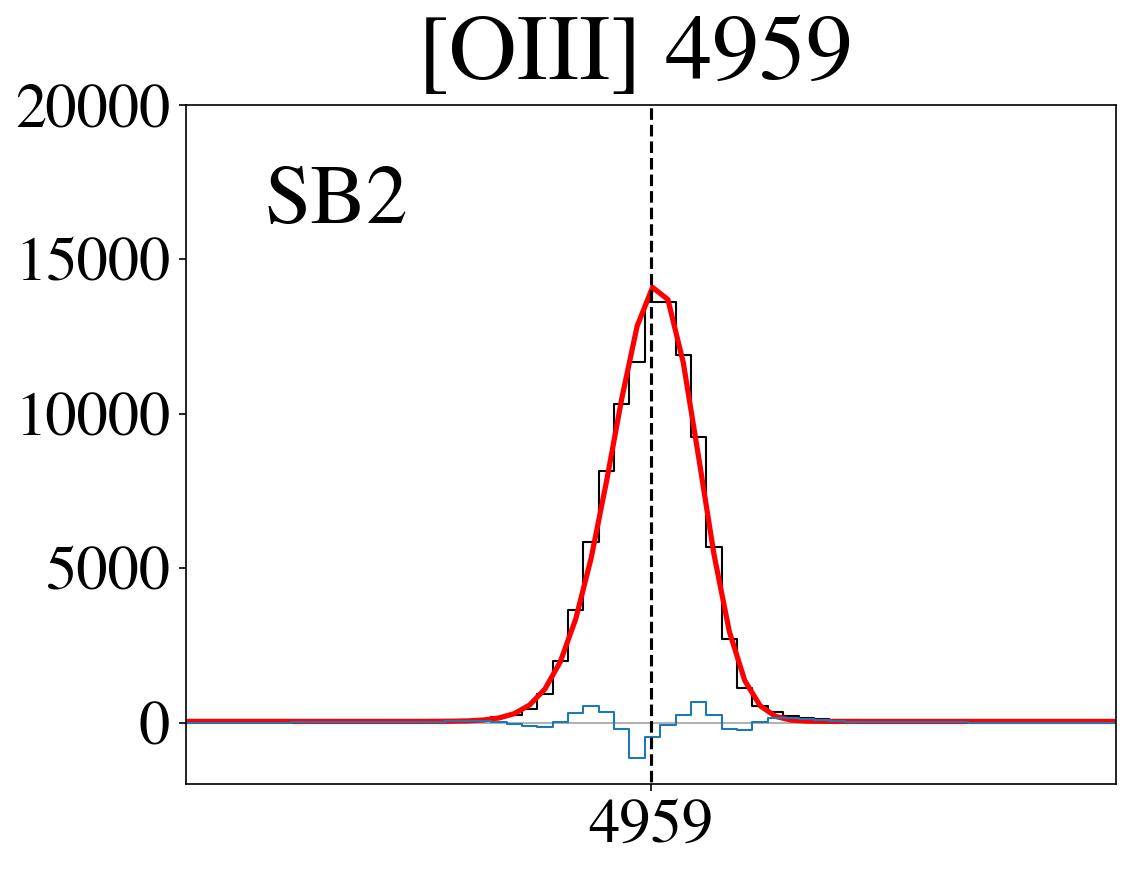}
    \vskip\baselineskip
    \includegraphics[width=1.75in]{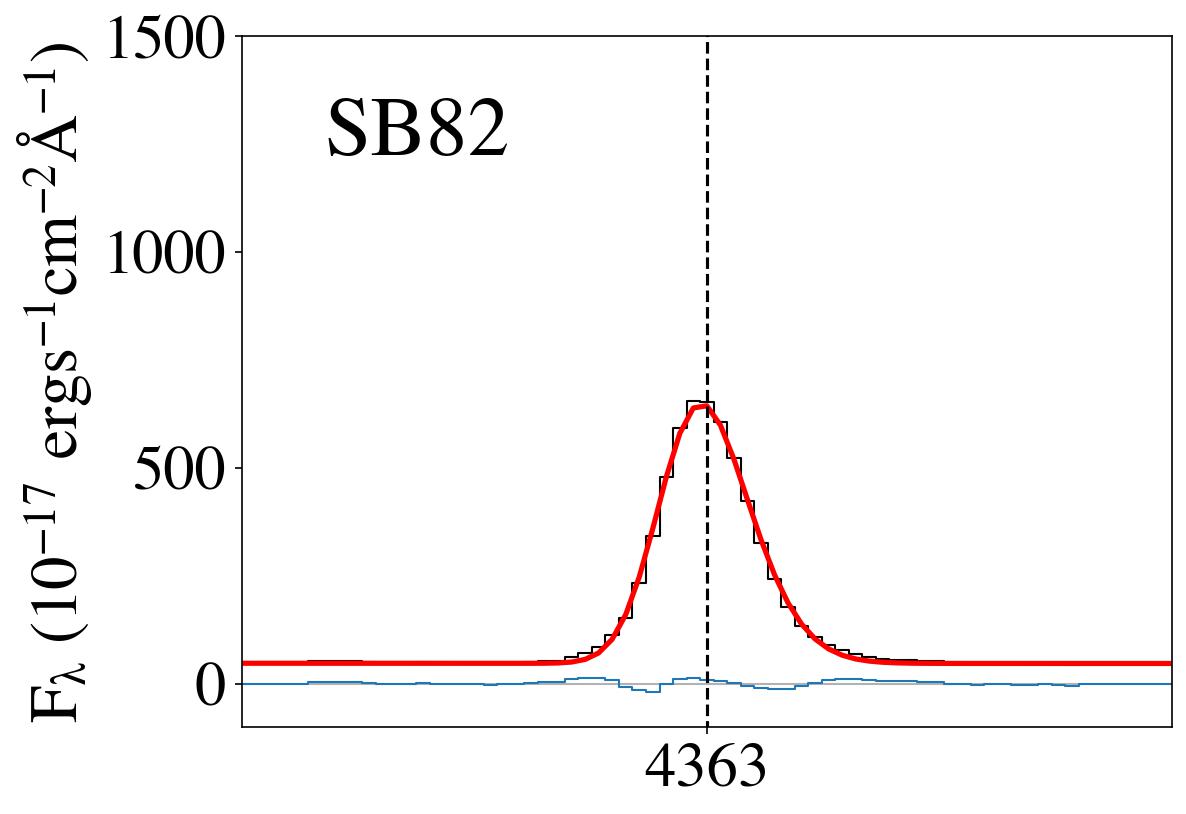} 
    \includegraphics[width=1.65in]{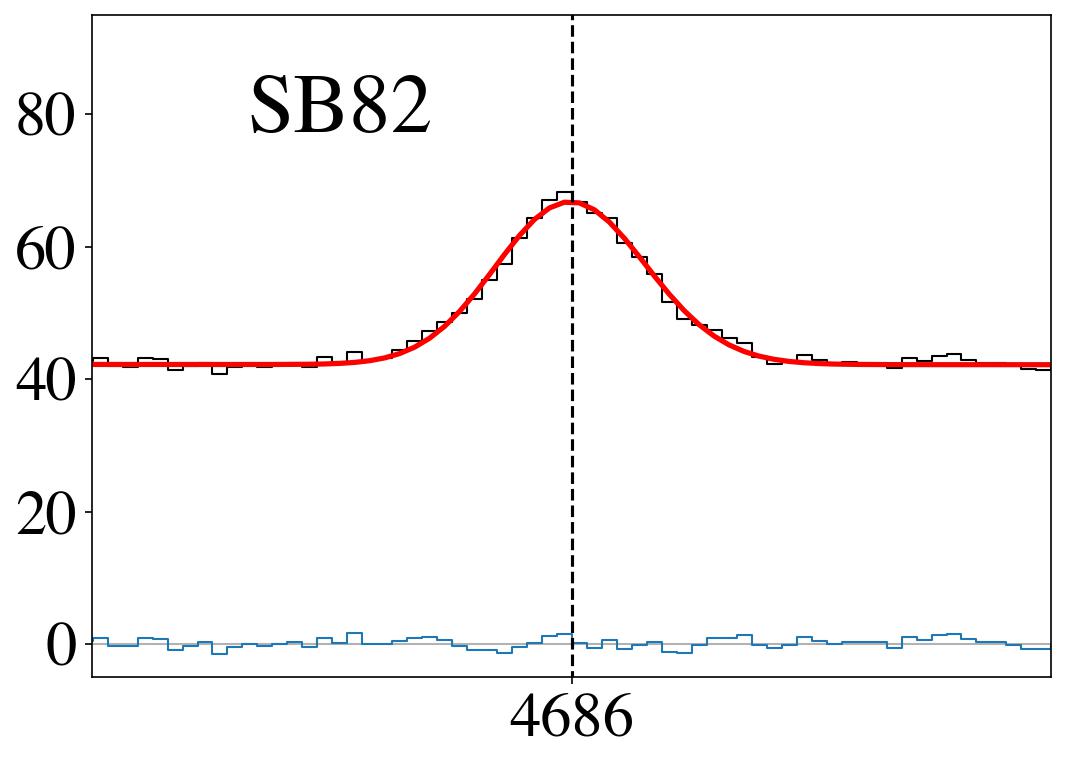}
    \includegraphics[width=1.65in]{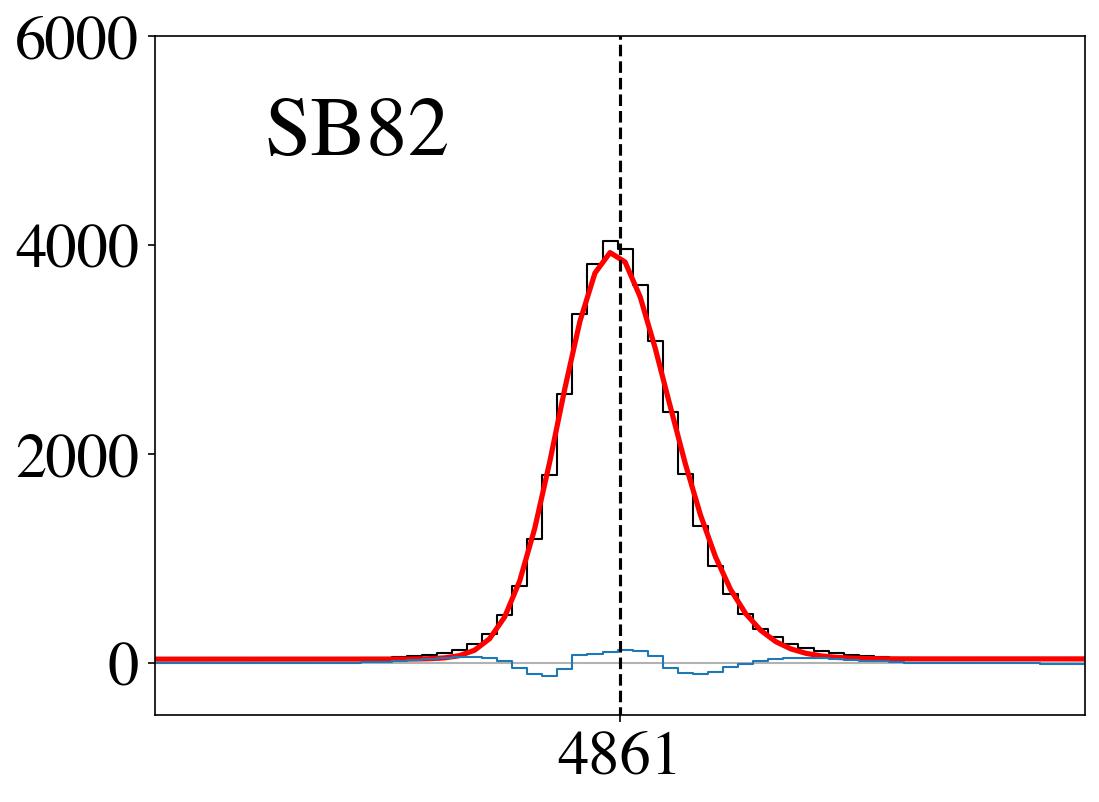}
    \includegraphics[width=1.65in]{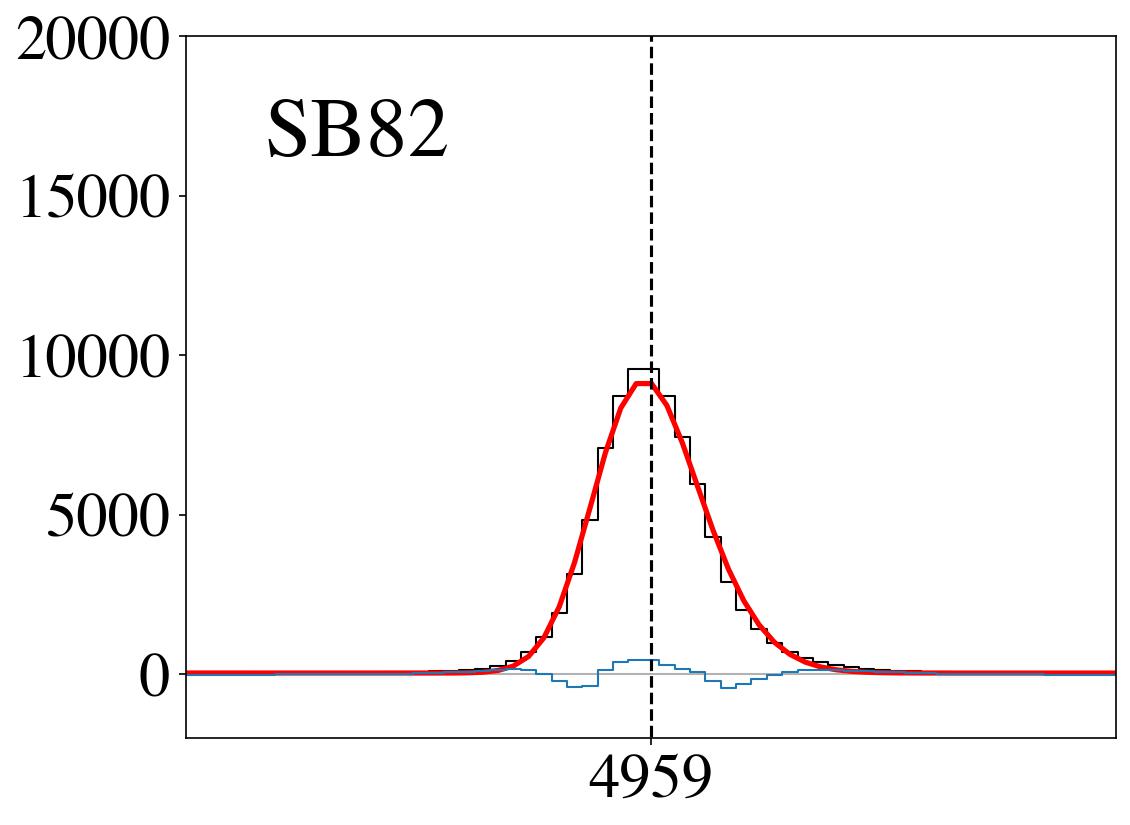}
    \vskip\baselineskip
    \includegraphics[width=1.75in]{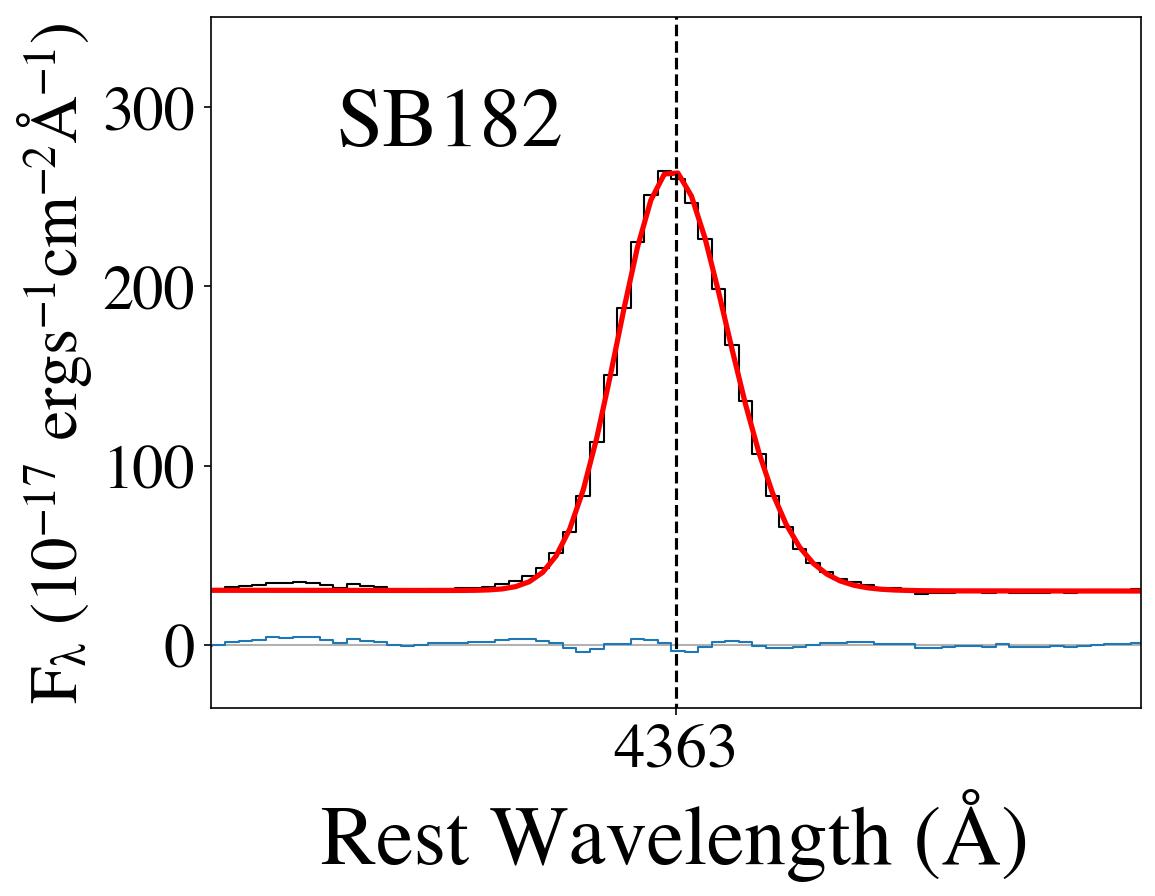} 
    \includegraphics[width=1.65in]{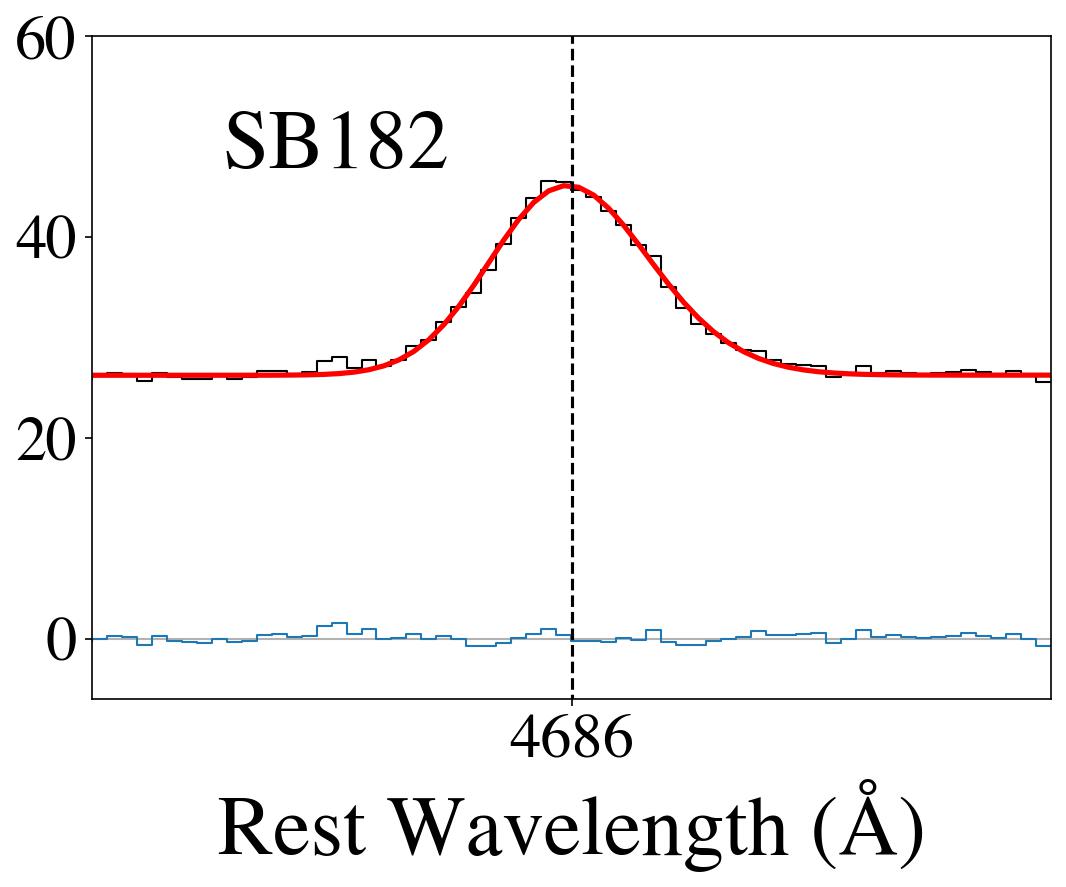}
    \includegraphics[width=1.65in]{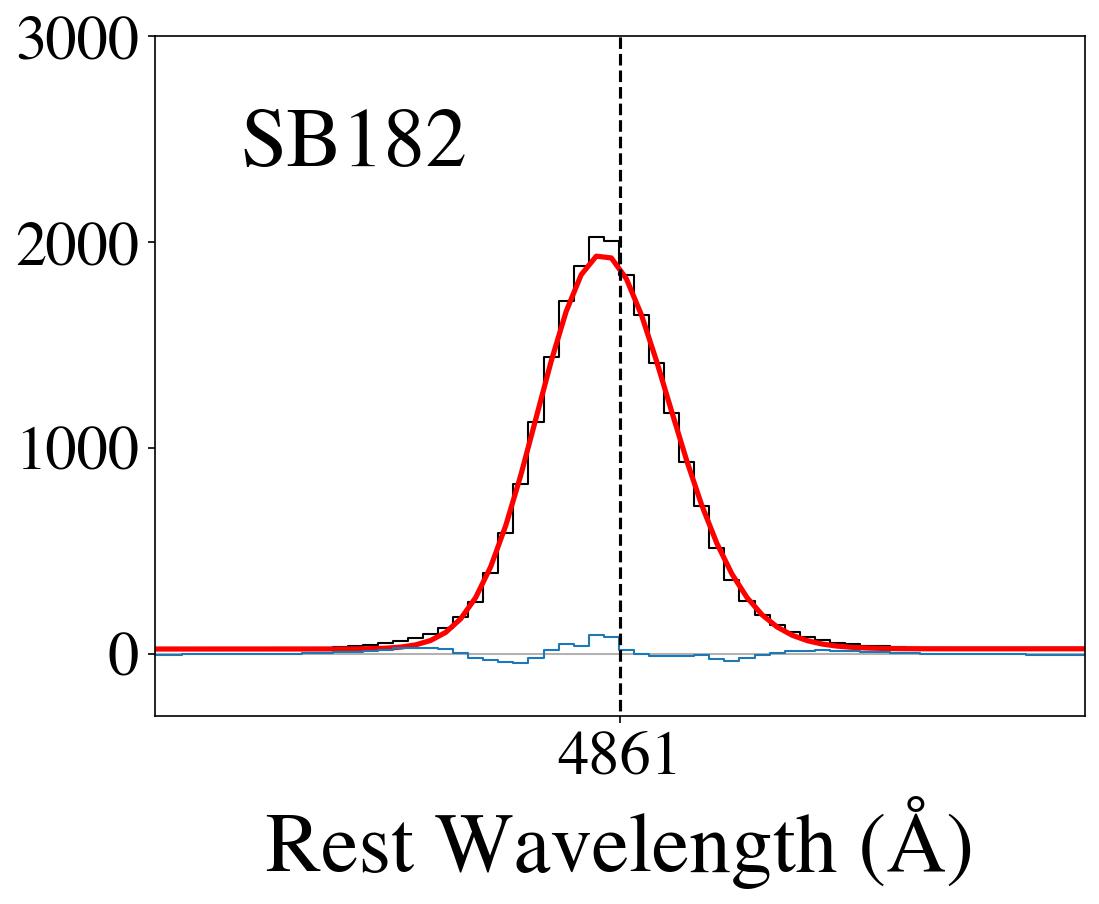}
    \includegraphics[width=1.65in]{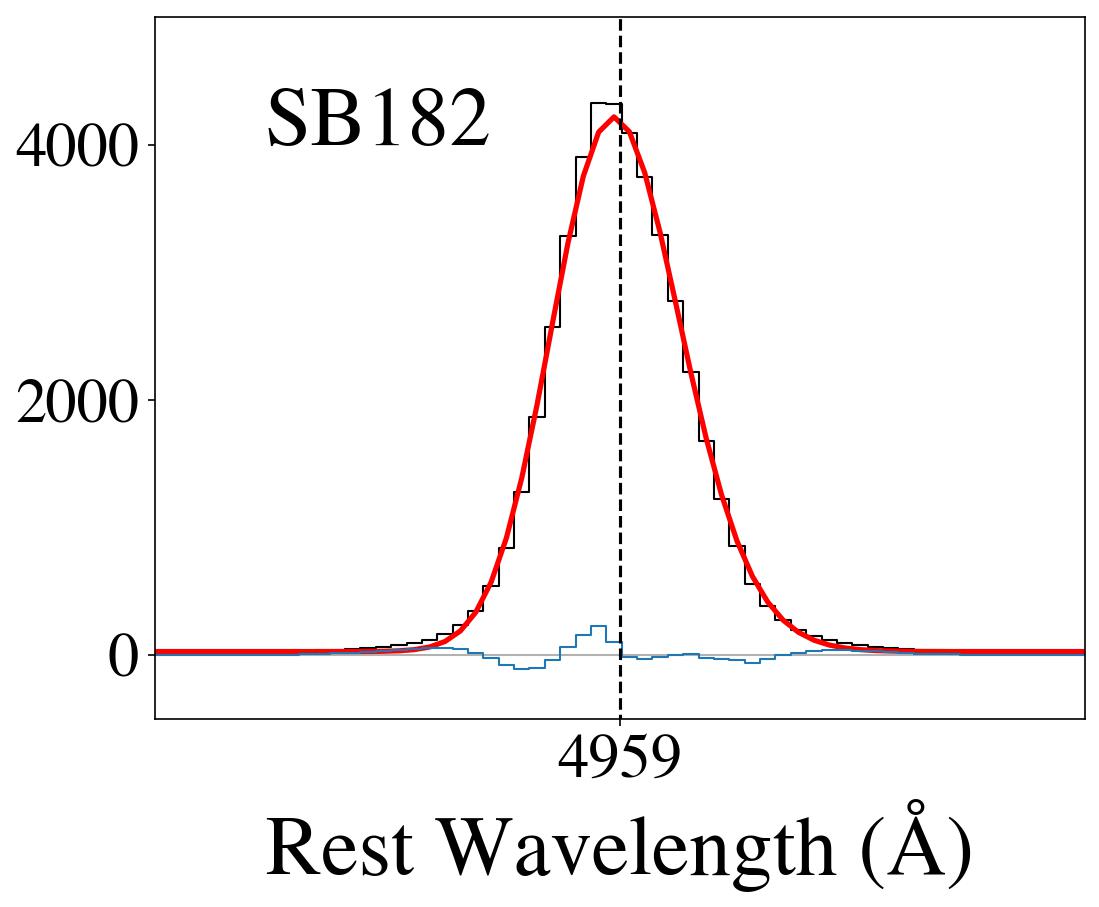}
    \caption{Rest-frame Keck/ESI spectra with the best-fit Gaussian profile and residual for select emission lines used in this analysis for each target. 
    The x-axis range is 10~\AA\ 
    for all spectra presented in this figure and the y-axis is at a scale of $10^{-17}$ erg\,s$^{-1}$\,cm$^{-2}$\,\AA$^{-1}$. We adopt a double Gaussian profile for \Heiio\ in SB~82, representing both the broad stellar and narrow nebular components. The broad component spans $\gtrsim~100$~\AA\ so it is not shown at the scale of this figure, but can be seen in Figure \ref{fig:opt_flux_fits_full}. All other lines are adequately fit with a single narrow component.}
    \label{fig:opt_flux_fits}
\end{figure*}

{While a large number of emission lines are detected in the Keck/ESI spectra, our analysis requires only a modest subset {detected at very high signal-to-noise ratios}: the \Hi\ Balmer and Paschen series, \Oiiit, \Oiiin, \Heiio, and the density-sensitive \Siios, 6731. We measured line fluxes by fitting Gaussian profiles superposed on a linear continuum.}
\Heiio\ can have a broad component from winds of massive stars as well as a narrow nebular component, of which only the nebular component is of interest to this analysis.
We thus include both a broad and narrow Gaussian component when fitting \Heiio, but only report fluxes for the narrow nebular component.
SB~2 and SB~182 display no broad component in the ESI spectra.
SB~82 shows weak but detected broad \Heiio\ emission, with the peak flux density of the broad component $\approx5\times$ smaller than the narrow peak flux density.
Visual inspection of the best fits indicates that the two-Gaussian model adequately matches the observed \Heiio\ profile in SB~82 and reliably recovers the flux of the narrow component.
Table~\ref{tab:flux} gives the observed optical line fluxes relative to H$\beta$, in units where I(H$\beta$)=100. We show best-fit profiles to the most relevant features (namely {\Oiiit}, {\Heiio}, {\Hb}, and {\Oiiin}) in Figure~\ref{fig:opt_flux_fits}.

We do not use the \Ha\ or \Oiiis\ lines because their peaks are nonlinear or saturated in the ESI spectra. Including these lines would not significantly affect the results, as equivalent information is obtained from \Oiiin\ and numerous \Hi\ lines (most importantly \Hb\ and \Hg). We have confirmed that the optical line fluxes used in this analysis are consistent between the SDSS and ESI spectra. In this work we use the ESI data which has {significantly} higher signal-to-noise and spectral resolution.

\subsubsection{UV Line Fitting}

The {\Heiiu}, {\Oiiiuo}, and {\Oiiiu} UV emission lines were originally described in \citetalias{Senchyna2017}, based on HST/COS spectra with limited exposure times (one orbit) for all three objects. Subsequent observations with significantly longer exposure times were obtained for SB2 and SB82, reported in \citetalias{Senchyna2022}. In this work we refit the UV line fluxes using the deeper spectra from these new observations. 
All three lines were fit using a Gaussian profile plus a linear continuum. {Milky Way interstellar absorption features were masked out where necessary}. The {\Oiiiu} and {\Oiiiuo} lines originate from the same upper energy level, resulting in a fixed {\Oiiiu}/{\Oiiiuo} flux ratio that is independent of $T_e$ and $n_e$. Thus, we performed a joint fit to {\Oiiib~$\lambda\lambda$1661,1666} with the same width and redshift for both lines, and fixing the flux ratio to the theoretically expected value of 2.49 which we calculated using \textsc{Pyneb} \citep{Luridiana2015}. 

\begin{figure*}
    \centering
    \includegraphics[width=3.in]{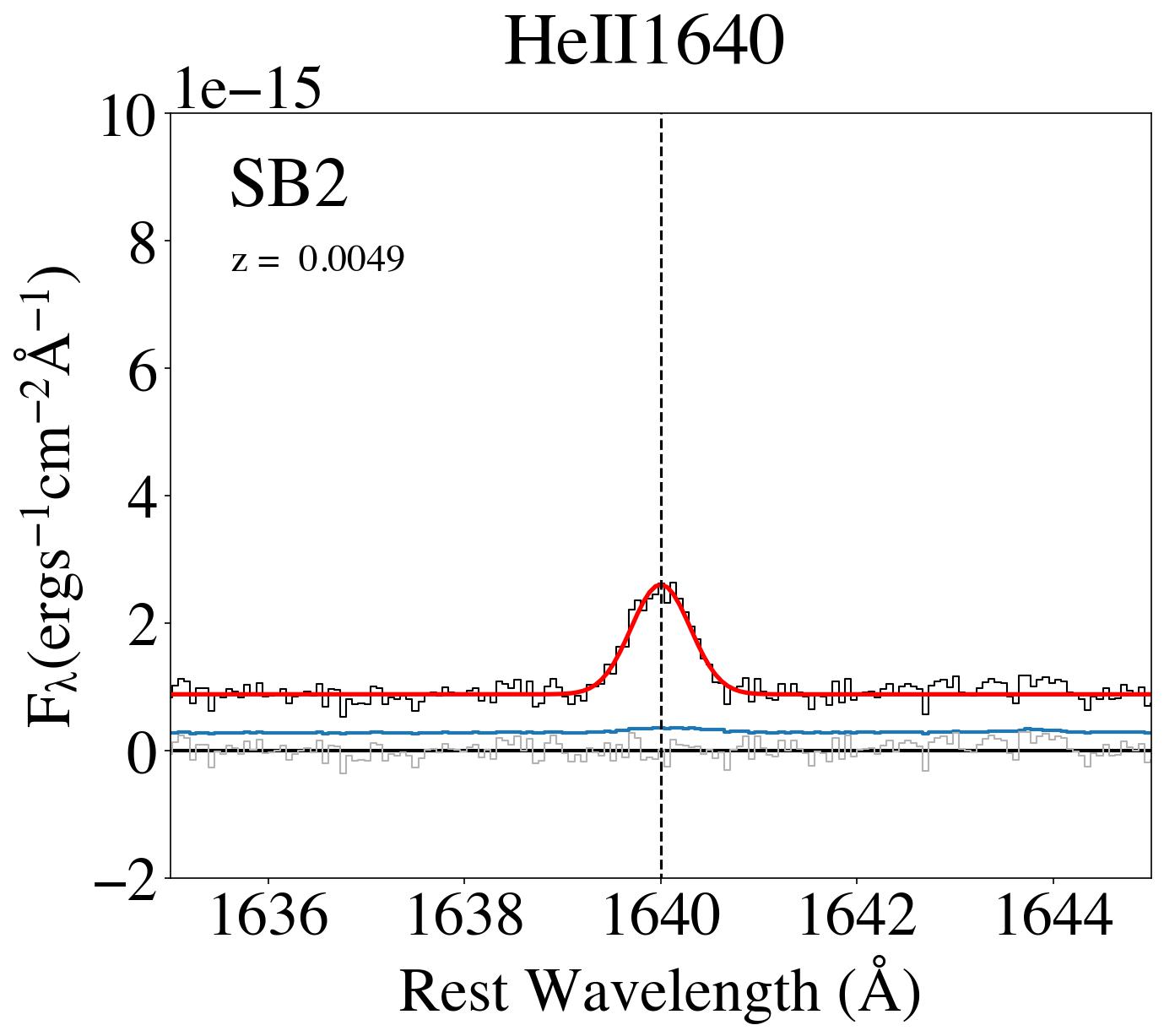}
    \includegraphics[width=3.85in]{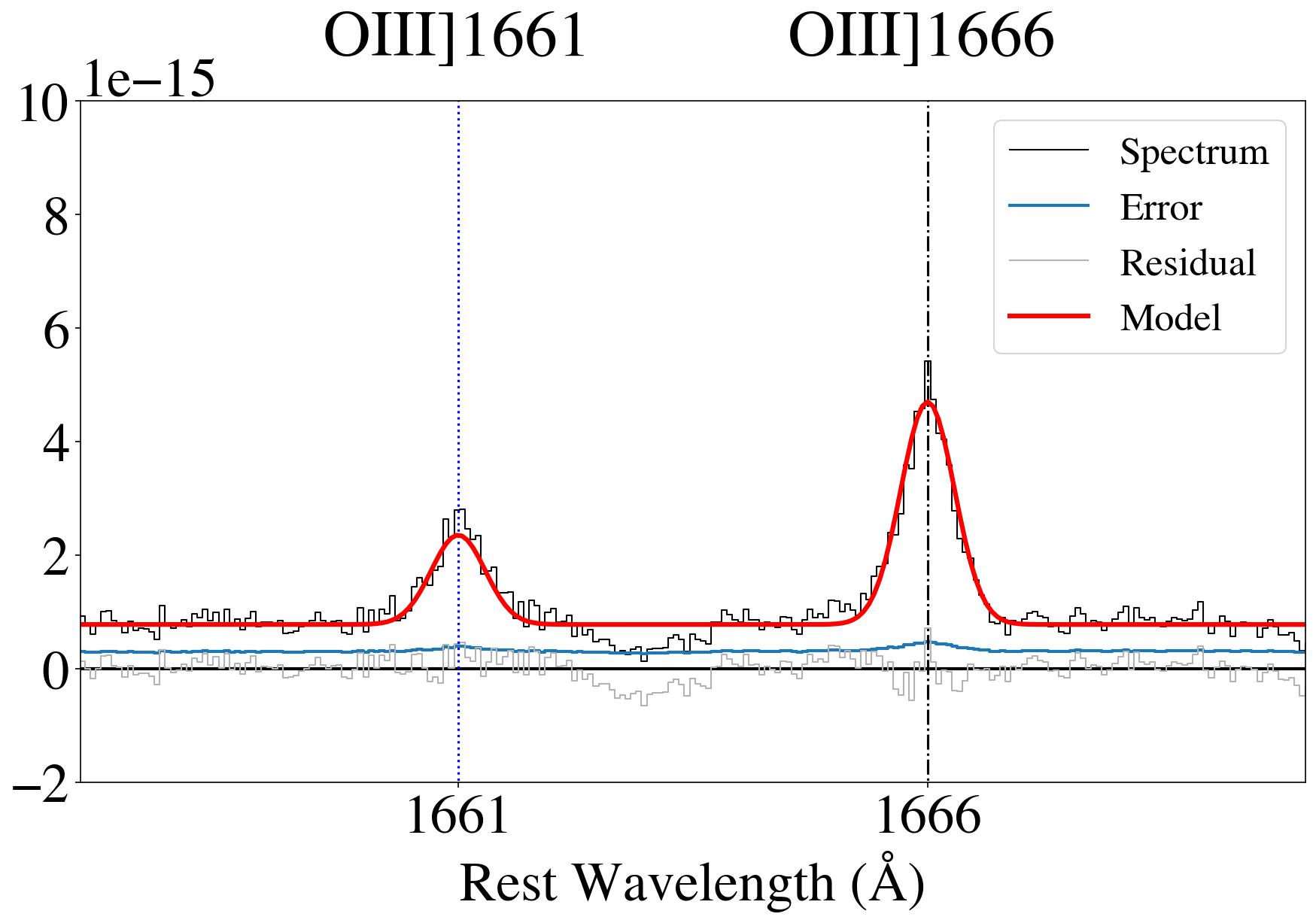}
    \vskip\baselineskip
    \includegraphics[width=3.in]{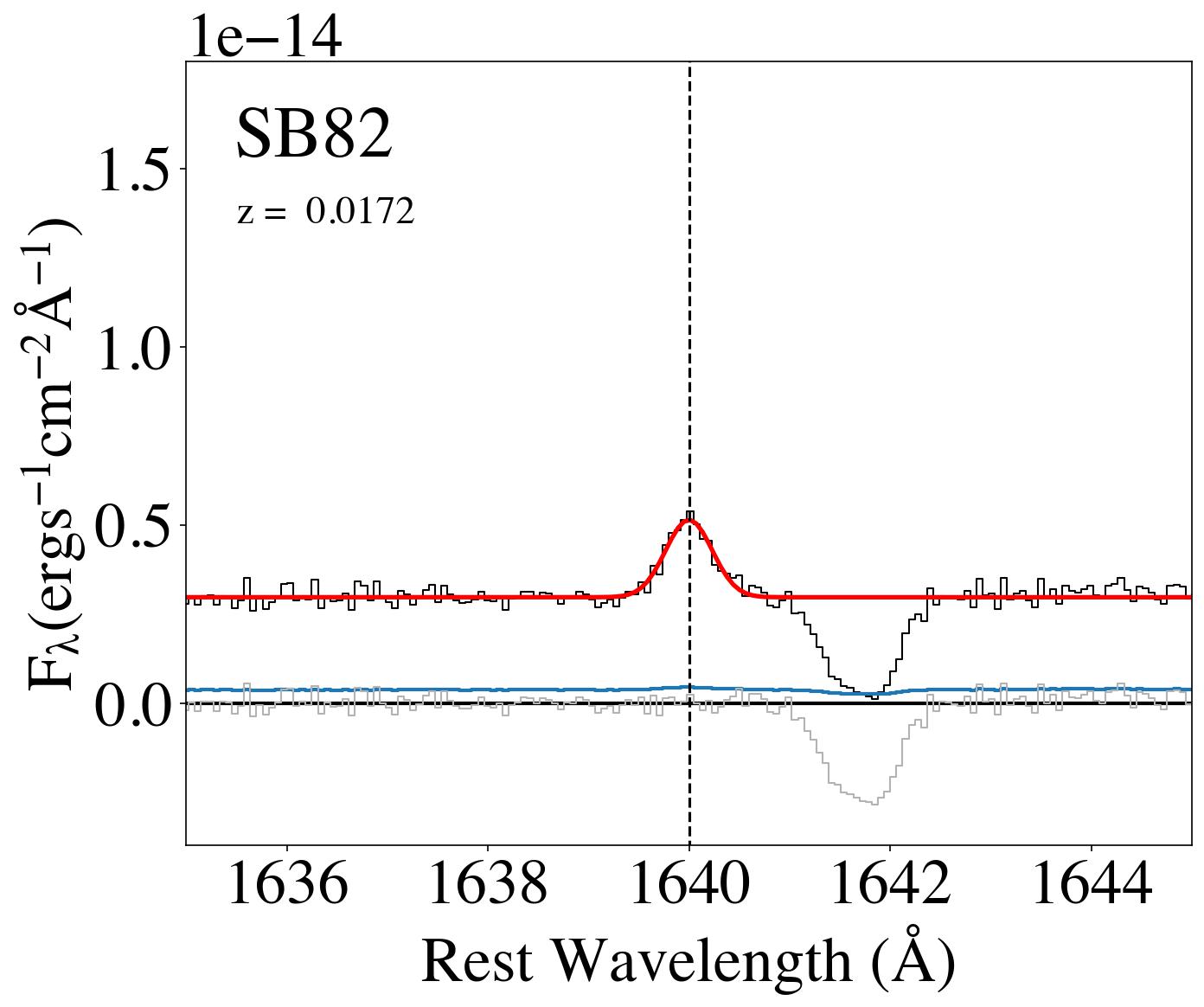}
    \includegraphics[width=3.85in]{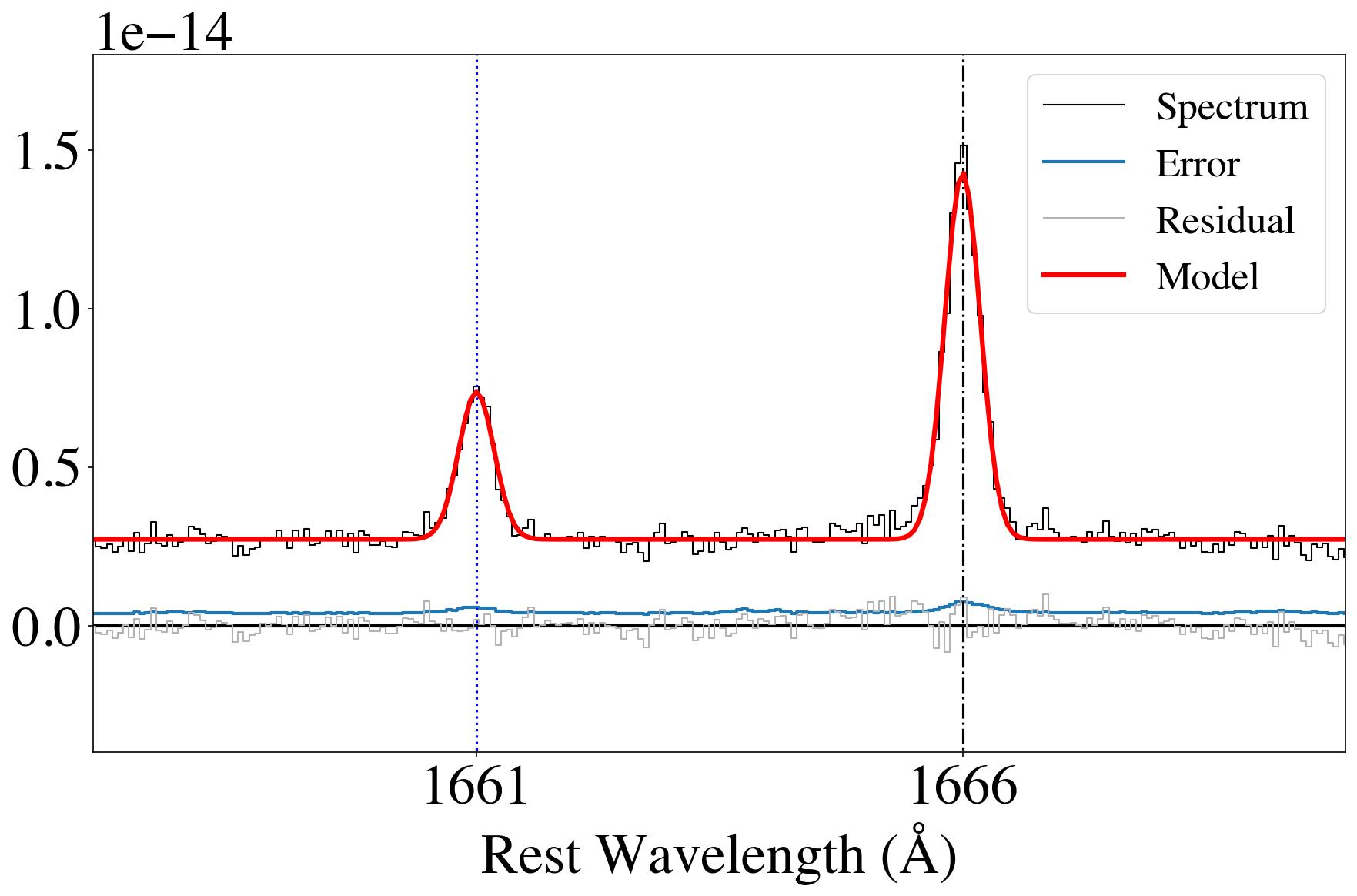}
    \vskip\baselineskip
    \includegraphics[width=3.in]{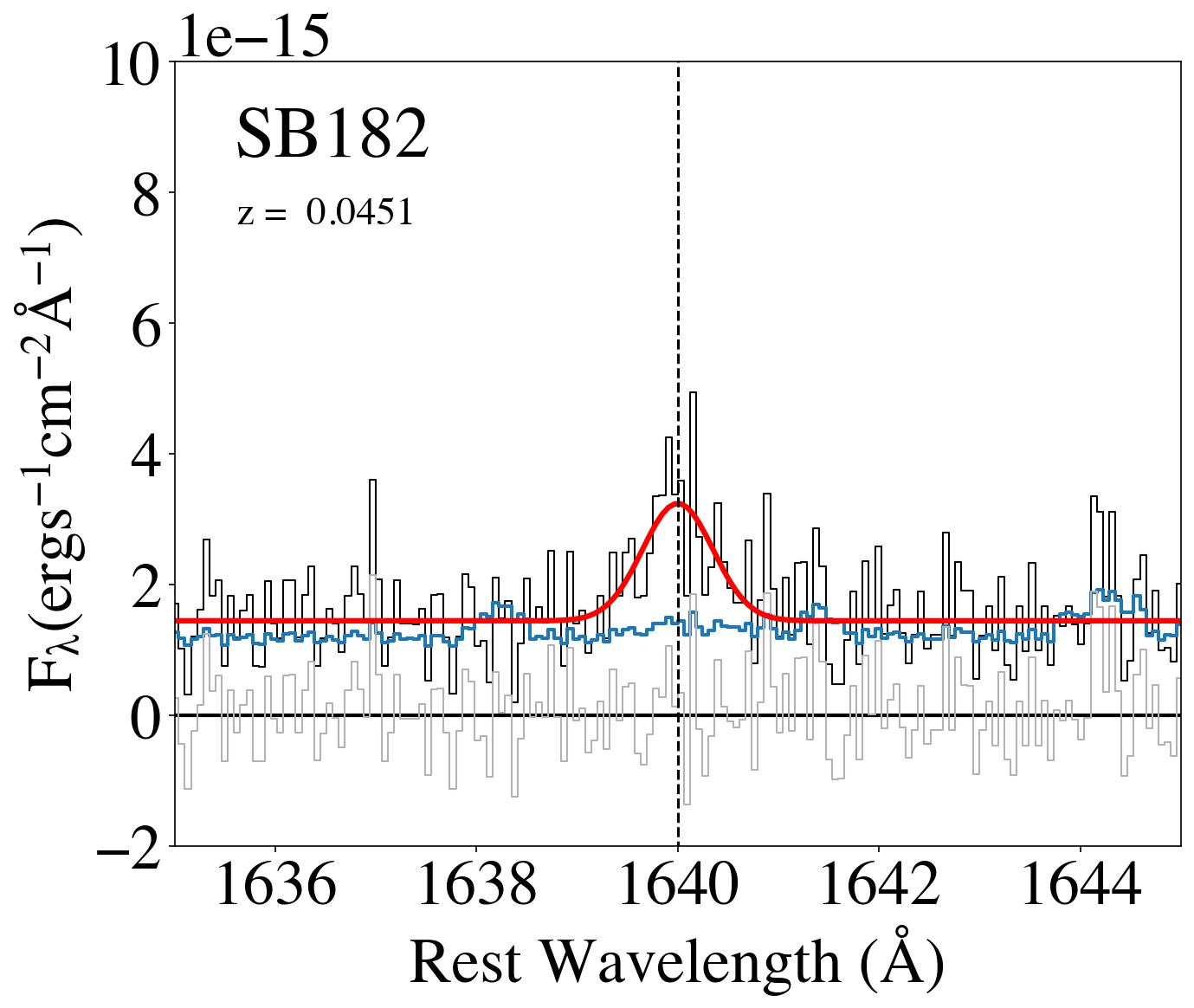}
    \includegraphics[width=3.85in]{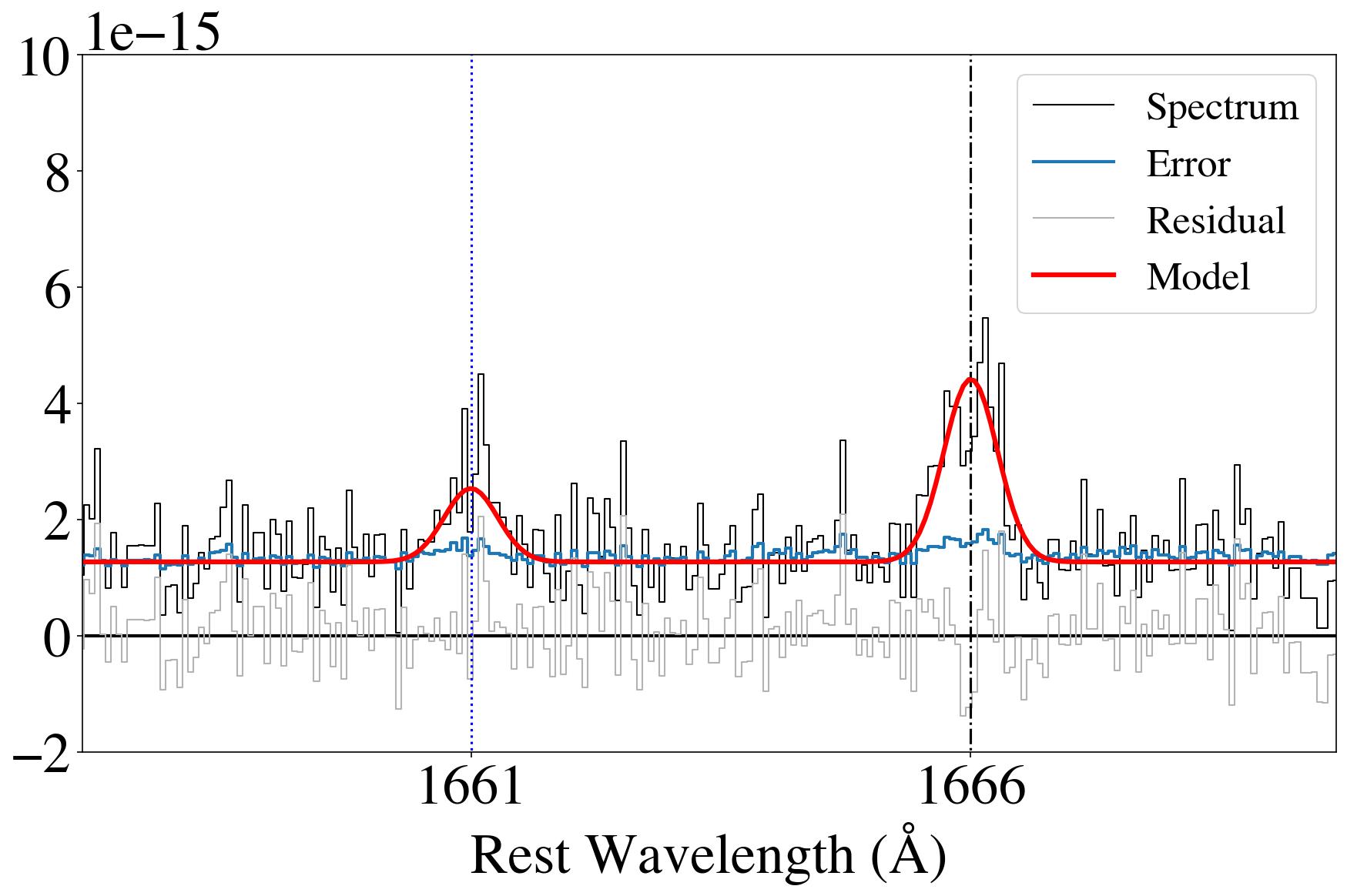}
    \caption{ Binned HST/COS spectra, associated 1$\sigma$ error spectra, best-fit Gaussian models, and residual spectra for our targets SB~2, SB~82 and SB~182. SB~2 and SB~82 have 10 orbits of COS data \citep{Senchyna2022}, while SB~182 has only a single orbit \citep{Senchyna2017}. Above, $z$ is the redshift as measured from the observed UV spectra. $f_{1640}$, $f_{1661}$ and $f_{1666}$ are the measured flux and {1$\sigma$ uncertainty} for {\Heiiu}, {\Oiiiuo} and {\Oiiiu} respectively. 
    Milky Way {interstellar} absorption features {(e.g., at rest wavelengths of approximately 1663~{\AA} for SB~2 and 1642~{\AA} for SB~82) were masked and excluded from the fitting. All emission lines used in our analysis are well-detected and unaffected by Milky Way absorption}.}
    \label{fig:uv_flux_fits}
\end{figure*}

The {\Heiiu} emission was fit independently from the {\Oiiib~$\lambda\lambda$1661,1666} doublet. The redshifts measured from \Heii\ and \Oiiib\ are consistent with one another for each object.
While SB~82 exhibits a broad component of optical \Heiio, we do not detect significant broad \Heiiu\ emission.
Given the flux ratio between the broad and narrow components of the optical \Heiio\ emission, we indeed do not expect to detect the broad component in the \Heiiu\ observations (see Section \ref{sec:opt_line_fit} and Figures \ref{fig:opt_flux_fits_full}, \ref{fig:opt_flux_fits}).
Likewise we do not detect broad \Heiiu\ in SB~2 and SB~182. Consequently we fit the UV \Heiiu\ lines with a single narrow component, which provides an adequate fit to the data. The resulting nebular emission fluxes are reported in Table~\ref{tab:flux}, and lines fits are shown in Figure~\ref{fig:uv_flux_fits}.

\subsection{Dust and Aperture Corrections}
\label{sec:reddening}

Accurately measuring metallicities from UV emission lines requires reliable comparisons between the UV and optical fluxes. A key objective of this work is to evaluate the effectiveness of using \Heii\ recombination lines for dust and aperture corrections. In this section, we outline the methodology to perform these corrections.

\subsubsection{Optical Reddening Correction}
\label{sec:OptReddening}

Before applying the dust and aperture corrections between the optical and UV, we first corrected the optical line flux ratios for dust attenuation. 
This correction was performed using the standard nebular reddening correction method based on \Hi\ Balmer and Paschen line ratios.
First, multiple $E(B-V)$ values were calculated using the ratios of {\Hg}, {\Hd}, and the 8750.47, 8862.79, 9014.91 and 9229.01 \AA\ Paschen lines (P12, P11, P10, P9) relative to {\Hb}. {Our fiducial analysis assumes the \citet{Cardelli1989} extinction law with $R_V = 3.1$, while alternative extinction laws and $R_V$ values are discussed below in this section}.
Intrinsic flux ratios relative to {\Hb} were calculated in \textsc{PyNeb} assuming an electron temperature $T_e =$~15,000 K and an electron density $n_e =$~175 cm$^{-3}$, corresponding to typical values in our sample (Section~\ref{sec:Te}).
The final E(B-V) value was then calculated as the inverse-variance weighted mean of the individual $E(B-V)$ estimates.
{\Ha} was not used due to saturation. The dust corrected optical fluxes normalized to I~({\Hb}) = 100 are listed in Table~\ref{tab:flux}.
While we report fluxes normalized to \Hb\ as is standard practice, \Heiio\ is the most relevant reference point for the joint analysis of optical and UV lines.

In order to account for systematic uncertainties arising from optical dust attenuation correction, we considered several extinction laws prior to choosing the well-established \cite{Cardelli1989} curve for our analysis. Additionally, we use a range of H-Balmer and Paschen lines for the correction, which cover the spectral range of $\sim$ 4100 -- 9200~\AA. The relative reddening correction between \Oiiin\ and {\Oiiit}, which both fall well within the range covered by the hydrogen lines, should not be sensitive to the choice of reddening curve. We tested three curves \citep{Cardelli1989, Calzetti2000, Gordon2003} with varying $R_v = 3.1$--$4.1$. The change in the \Oiiit/\Oiiin\ ratio is within 0.7\% of our fiducial choice.

\subsubsection{UV Reddening and Aperture Correction}
\label{sec:UVreddening}

Dust attenuation laws exhibit significant variation in the ultraviolet \citep[e.g., ][]{Cardelli1989,Calzetti2000,Gordon2003}, introducing substantial systematic uncertainties in UV-to-optical line ratios (e.g., {\Oiiiuvratio}). Furthermore, differences in the acquisition apertures and point-spread functions between the Keck/ESI and HST/COS spectra complicate direct comparisons. By leveraging the UV {\Heiiu} and optical {\Heiio} recombination lines, both issues are simultaneously addressed. 

The intrinsic flux ratio {\Heiiu}/{\Heiio} $= 6.99$ ({\Heiio}/{\Heiiu} $= 0.143$) at $T_e = 15000$~K and $n_e = 175$~cm$^{-1}$, and is relatively insensitive to variations in $T_e$ and $n_e$ over the range of temperatures we investigate, {making this ratio a reliable tool for correcting dust reddening between UV and optical nebular lines}.
This approach is similar to the use of \Hi\ line ratios in correcting optical attenuation. (We note that while the {\Heiio}/{\Heiiu} ratio is relatively stable, it is more affected by $T_e$ than {\Hi}. We address this concern later in this section.)
After the optical lines have been corrected for {dust} reddening (as described in Section~\ref{sec:OptReddening}), the reddening-corrected and aperture-matched {\Heiiu} flux can be determined by comparing the observed {\Heiio}/{\Heiiu} flux ratio to its intrinsic value.
The correction factor $f_{c}$ can be expressed as
\begin{equation} \label{eq:UVcorr}
      {f_c} = \left({\mathrm{\frac{He~{\sc{II}}~4686_{~corr~}}{He~{\sc{II}}~1640_{~unc}}}}\right)_\mathrm{obs} \times~ \left({\mathrm{\frac{  He~{\sc{II}}~1640}{He~{\sc{II}}~4686}}}\right)_\mathrm{int},
\end{equation}
\normalsize
where {\Heiio}$_\mathrm{corr}$ is the dust-corrected {\Heiio} flux, {\Heiiu}$_\mathrm{unc}$ is the uncorrected observed {\Heiiu} flux, and ({\Heiiu}~/~{\Heiio})$_\mathrm{int}$ is the intrinsic flux ratio between the two lines. 

Once this correction factor is determined, it can be applied to the {\Oiiiu} and {\Oiiiuo} lines. Since the {\Oiiiu}, {\Oiiiuo} doublet and {\Heiiu} are relatively close in wavelength, the variation in the correction factor between these lines is minimal.
For the \citet{Cardelli1989} extinction curve with $R_V = 3.1$ and $E(B-V) = 0.167$ {(the maximum E(B-V) value among the three objects)}, the correction factor changes by only 0.5\% between {\Oiiiu} and {\Heiiu}. 
For the {\citet{Calzetti2000} curve with $R_V = 3.1$, the difference in correction factor is only 1\%} and for the \citet{Gordon2003} curve with $R_V = 3.41$ {(2.74)}, the difference is only 0.2\% {(0.15\%)}.
This difference is well within the uncertainties of our analysis, regardless of the chosen attenuation curve. Thus we apply the \Heii-based correction factor to the UV \Oiiib~$\lambda\lambda$1661,1666 emission lines via
\begin{equation}
    { ~~\mathrm{F_{\lambda i~(corr)} = {F_{\lambda i~(unc)}}~ \times~} {f_c}},
\end{equation}
where F$\mathrm{_{\lambda i~(corr)}}$ and F$\mathrm{_{\lambda i~(unc)}}$ are then the corrected and uncorrected fluxes for transition $i$ respectively. The corrected UV line fluxes are reported in units of I({\Hb}) = 100 in Table~\ref{tab:flux}. 

While the intrinsic {\Heiio}/{\Heiiu} flux ratio is relatively stable, it does exhibit a slight dependence on $T_e$, which could introduce systematic biases into our analysis. The relationship between $T_e$ in different ionic gas regions, including between He$^+$ and \Opp\ zones, remains an active topic of investigation \citep[e.g.,][]{Esteban2014, kreckel22, rickards-vaught24}. Furthermore, {nebular zones with different degrees of ionization may be correlated with different levels of temperatue fluctuations} \citep{mendez-delgado22}.
To account for this, we incorporate a systematic uncertainty of $\pm$ 4,000 K in the \Heii\ $T_e$ when deriving the correction factor. This choice is motivated by the typical $T_e$ differences observed between low-ionization (e.g., \Oii) and high-ionization (e.g., \Oiii) zones in low-metallicity, high-ionization \Hii\ regions \citep[e.g.,][]{rickards-vaught24}. This uncertainty is added in quadrature to the UV emission line flux measurement uncertainties. 
The weight that this uncertainty carries in our final results varies by object. Error on {the quantification of temperature fluctuations in our final results (see \S\ref{sec:t2}: \emph{Temperature Fluctuations})} for SB~2 and SB~182 is not substantially affected by the uncertainty introduced by the intrinsic {\Heiio}/{\Heiiu} ratio 
because uncertainty from flux measurements dominates. However, for SB~82, the error introduced by the temperature dependence of the correction factor is co-dominant with the UV flux error (i.e., signal-to-noise) and accounts for approximately 30\%\ of the $t^2$ error budget. 

\begin{table*}
\centering
	\caption{Observed and de-reddened fluxes in units of I~({\Hb}) = 100. }
	\label{tab:flux}
	\begin{tabular}{lcccccc}
	\hline
		{\bf Object:}  &  \multicolumn{2}{c}{SB~2} & \multicolumn{2}{c}{SB~82} & \multicolumn{2}{c}{SB~182}
		\vspace{1.mm} \\
        Transition &  Uncorrected & Corrected & Uncorrected & Corrected &  Uncorrected & Corrected \\
		\hline
        \multicolumn{7}{l}{\textbf{Optical (Keck~/~ESI):}} \\
        {\Hd}  & 24.1  $\pm$ 0.3 & 27.3  $\pm$ 1.1 & 24.3  $\pm$ 0.3 & 24.7  $\pm$ 1.1 & 26.3  $\pm$ 0.3 & 26.3  $\pm$ 0.8 \\[2.pt]	
		{\Hg}{$^a$}  & 44.2  $\pm$ 1.0 & 48.2  $\pm$ 2.2 & 45.7  $\pm$ 0.8 & 46.3  $\pm$ 2.2 & — & — \\[2.pt]
		{\Oiiit}  & 11.8  $\pm$ 0.1 & 12.8  $\pm$ 0.5 & 12.8  $\pm$ 0.1 & 13.0  $\pm$ 0.6 & 10.2  $\pm$ 0.1 & 10.2 $\pm$ 0.3 \\[2.pt]	
		{\Heiio}  & 1.09  $\pm$ 0.04 & 1.12  $\pm$ 0.06 & 0.661  $\pm$ 0.022 & 0.664  $\pm$ 0.037 & 0.972  $\pm$ 0.027 & 0.972  $\pm$ 0.038 \\[2.pt]	
		{\Hb}  & 100.0  $\pm$ 1.7 & 100.0  $\pm$ 4.3 & 100.0  $\pm$ 1.4 & 100.0  $\pm$ 4.7 & 100.0  $\pm$ 1.5 & 100.0  $\pm$ 3.1 \\[2.pt]
		{\Oiiin}  & 187.  $\pm$ 3. & 185.  $\pm$ 8. & 229.  $\pm$ 3. & 228.  $\pm$ 11. & 213.  $\pm$ 2. & 213.  $\pm$ 6. \\[2.pt]
		{\Sii}~6716  & 9.38  $\pm$ 0.15 & 7.86  $\pm$ 0.33 & 8.27  $\pm$ 0.09 & 8.05  $\pm$ 0.37 & 11.6  $\pm$ 0.2 & 11.6  $\pm$ 0.4 \\[2.pt]
		{\Sii}~6731  & 7.19  $\pm$ 0.08 & 6.02  $\pm$ 0.24 & 6.44  $\pm$ 0.04 & 6.27  $\pm$ 0.28 & 8.99  $\pm$ 0.09 & 8.98  $\pm$ 0.26 \\[2.pt]
		P12  & 1.34  $\pm$ 0.02 & 0.980  $\pm$ 0.040 & 1.08  $\pm$ 0.02 & 1.02  $\pm$ 0.05 & 0.950  $\pm$ 0.016 & 0.948  $\pm$ 0.031 \\[2.pt]
		P11  & 2.06  $\pm$ 0.04 & 1.49  $\pm$ 0.06 & 1.26  $\pm$ 0.02 & 1.20  $\pm$ 0.06 & 1.33  $\pm$ 0.02 & 1.33  $\pm$ 0.04 \\[2.pt]
		P10  & 2.50  $\pm$ 0.04 & 1.80  $\pm$ 0.08 & 1.97  $\pm$ 0.02 & 1.87  $\pm$ 0.09 & 1.74  $\pm$ 0.08 & 1.74  $\pm$ 0.10 \\[2.pt]
		P9  & 3.39  $\pm$ 0.07 & 2.42  $\pm$ 0.11 & 2.53  $\pm$ 0.05 & 2.40  $\pm$ 0.12 & 2.20  $\pm$ 0.04 & 2.19  $\pm$ 0.07 \\[2.pt]
		\hline
        {\bf E(B-V):}  &  \multicolumn{2}{c}{0.167 $\pm$ 0.017} & \multicolumn{2}{c}{0.035 $\pm$ 0.016} & \multicolumn{2}{c}{0.001 $\pm$ 0.012} \\
		\hline
        {\it \Hb\ flux {$^c$}}: & \multicolumn{2}{c}{9007. $\pm$ 47.} & \multicolumn{2}{c}{5933. $\pm$ 46.} & \multicolumn{2}{c}{3453. $\pm$ 25.}\\
        \hline
		\hline
		\multicolumn{7}{l}{\textbf{UV Reddening and Aperture Uncorrected and Corrected Lines (HST~/~COS):}}\\
		{\Heiiu} & 1.30 $\pm$ 0.04 & 7.77 $\pm$ 0.54 & 
		1.26 $\pm$ 0.08 & 4.50 $\pm$ 0.61 & 
		1.58 $\pm$ 0.23 & 6.75 $\pm$ 1.53 \\[2.pt]
		{\Oiiiuo}{$^b$} & 1.13 $\pm$ 0.02 & 6.77 $\pm$ 1.35 & 
		2.29 $\pm$ 0.03 & 8.14 $\pm$ 0.99 & 
		0.95 $\pm$ 0.08 & 4.06 $\pm$ 0.78 \\[2.pt]
		{\Oiiiu}{$^b$} & 2.81 $\pm$ 0.05 & 18.6 $\pm$ 1.97 &  
		5.70 $\pm$ 0.07 & 20.3 $\pm$ 2.5 & 
		2.37 $\pm$ 0.21 & 10.1 $\pm$ 1.9 \\
		\hline
	\end{tabular}
        \begin{footnotesize}
            \begin{flushleft}
                $^{\text{a}}$ {\Hg} in SB~182 is affected by a detector artifact, and was not used for optical reddening correction in this object.\\
                $^{\text{b}}$ From a joint fit of 
                \Oiiib~$\lambda\lambda$1661,1666.\\
                {$^{\text{c}}$ The observed \Hb\ fluxes before extinction correction (units: $10^{-17}\mathrm{erg~s^{-1}cm^{-2}\AA^{-1}}$).}
            \end{flushleft}
        \end{footnotesize}
\end{table*}

\subsection{Measurement of {\Opp} Electron Temperature}
\label{sec:Te}

To measure the \Opp\ abundance, the electron temperature ($T_e$) of the ionized gas must first be determined in order to calculate the emissivity of the \Oiii\ emission lines.
The $T_e$ can be derived from the ratio of collisionally excited lines {originating from different upper energy levels} \citep[e.g.,][]{osterbrock06}.
In this work we obtain two measurements of $T_e$ for each object using the dust-corrected flux ratios {\Oiiioptratio} and {\Oiiiuvratio}, hereafter referred to as {\Teopt} and {\Teuv} respectively. The $^{1}D_2 \rightarrow ^{3}P_1$ energy level transition produces the {\Oiiin} emission line, $^{1}S_0 \rightarrow ^{1}D_2$ produces {\Oiiit}, and $^{5}S_2^o \rightarrow ^{1}D_2$ produces {\Oiiiu}.
Each transition samples a different upper energy level for the \Opp\ ion, which consequently sample different collisional energies and are therefore sensitive to the electron temperature of the gas.

The $T_e$ calculations were performed in \textsc{Pyneb}. While this requires an input of the electron density ($n_e$), both the optical and UV \Oiii\ emission lines essentially have no dependence on $n_e$ across the typical range found in \Hii\ regions, since their critical densities for collisional de-excitation are $\gg 1000~\mathrm{cm}^{-3}$. For this work we adopt $n_e$ values derived from the {\densratio} line ratio, which yields $n_e$ = 158, 189, 184 cm$^{-3}$ for SB~2, SB~82, and SB~182 respectively (Table~\ref{tab:properties}).

We consider the possibility that the choice of atomic data may influence the measurements of $T_e$. This can in turn affect determinations of $t^2$ and $T_0$ (see Section~\ref{sec:t2}), as well as abundances measured using the $T_e$ method.
As such, measurements for $T_e$ and the $t^2$ analysis presented in this work were run using three different collisional strength data sets for {\Opp} \citep{Aggarwal1999,Tayal2017,Mao2021}.
We note that \cite{Mao2021} produced higher values for individual {\Teopt} and {\Teuv} measurements, but remain consistent within the uncertainties of the results obtained when adopting other collisional datasets. Additionally, the \emph{difference} between the two quantities was the same as those from other data sets, and as such, values for $t^2$ were not affected by the high individual temperature measurements. We conclude that the choice of atomic data does not significantly affect $t^2$ results and adopt \cite{Tayal2017} collisional strengths throughout this analysis.

\begin{figure}
    \centering
    \includegraphics[width=\columnwidth]{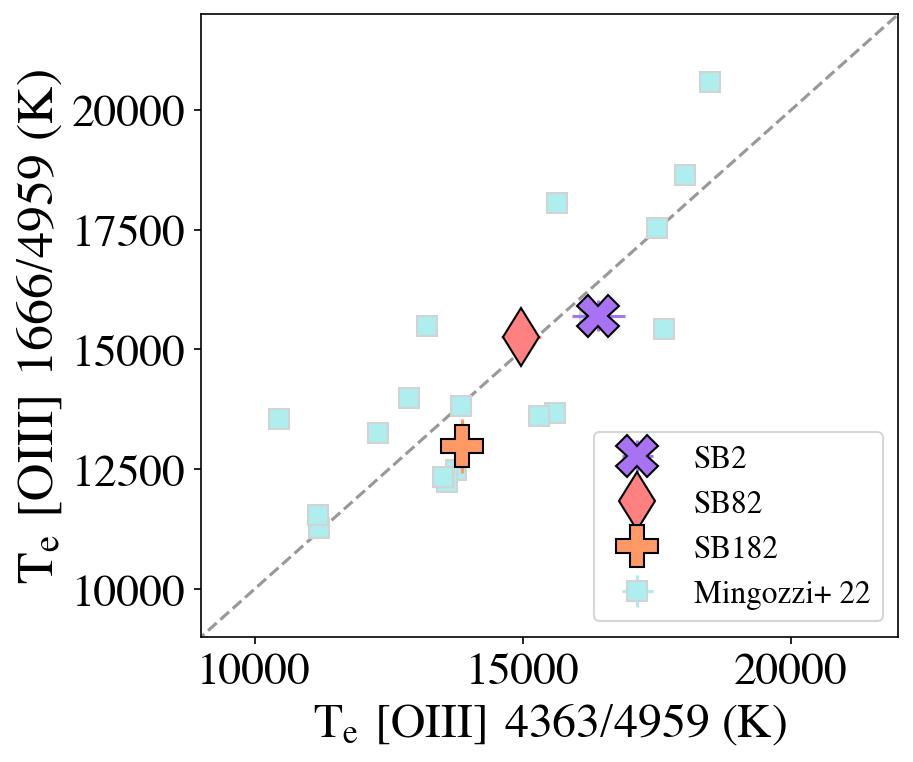}
    \caption{Our measurements of \Teuv\ versus \Teopt\ for the objects in our sample, along with measurements from \cite{Mingozzi_etal_2022} which used a different methodology. Two galaxies in our sample (SB~2 and SB~182) have \Teuv$<$\Teopt, whereas temperature fluctuations predict the opposite result.
    Our sample displays a tight distribution with standard deviation scatter of 515~K ($4.2\%$ mean absolute percent offset), while the \cite{Mingozzi_etal_2022} sample has a larger scatter of 1557~K ($9.2\%$ mean absolute percent offset).
    The improved precision demonstrates success in reducing the uncertainties associated with UV attenuation corrections using our \Heii\ method (see Section~\ref{sec:UVreddening}).}
    \label{fig:TePlot}
\end{figure}

The derived \Teopt\ and \Teuv\ are reported in Table~\ref{tab:Te}. Figure~\ref{fig:TePlot} compares their relation with similar measurements by \citet[][hereafter \citetalias{Mingozzi_etal_2022}]{Mingozzi_etal_2022}, who found rough agreement {between} the two values.
Our results produce a similar distribution along the 1:1 line. Although our sample contains only three galaxies, our measurements have smaller scatter about the 1:1 line compared to \citetalias{Mingozzi_etal_2022}. Our $T_e$ measurements have a standard deviation scatter of 515~K and a mean absolute percent offset of 4.2\%, while the points from \citetalias{Mingozzi_etal_2022} have a standard deviation scatter of 1557~K and a 9.2\% mean absolute percent offset. While this smaller scatter may arise in part from the small-sample statistics, or the narrower dynamic range of our sample properties, at face value it suggests that using \Heii\ emission lines for reddening and aperture corrections between the UV and optical provides more precise results.

\begin{table*}
    \centering
    \caption{Measured $T_e$, $T_0$, and $t^2$ values for SB~82, SB~182, and SB~2. We note that variances $t^2 < 0$ are formally unphysical. \\
    }
    \label{tab:Te}
    \begin{tabular}{l|c|cccc}
        \hline\hline
         {\bf Object}~ & ~$n_e$ (cm$^{-3}$) ~ & ~{\Teopt}~(K) & {\Teuv}~(K) & $T_0$~(K) & $t^2$\\
         \hline
         SB2~ &  158 & ~16410. $\pm$ 502. & 15703. $\pm$ 324. & 17672. $\pm$ 909. & -0.054 $\pm$  0.043\\
         SB82~ & 189 & ~14971. $\pm$ 373. & 15257. $\pm$ 365. & 14235. $\pm$ 950. &  0.024 $\pm$ 0.033\\
         SB182~ & 184 & ~13869. $\pm$ 259. & 12986. $\pm$ 536. & 15779. $\pm$ 702. & -0.078 $\pm$ 0.031\\
         \hline
    \end{tabular}
    
\end{table*}

\subsection{{Ionic} Oxygen Abundance}
\label{sec:abd}

Using the $T_e$ values calculated above {and reported in Table~\ref{tab:Te}}, we can derive the {\Opp}/{\Hp} abundance for each object by comparing the flux of \Oiii\ and \Hi\ lines {(namely \Oiiin/H$\beta$ which gives the most precise result)}. In this work, we provide three sets of oxygen abundances in the form of  12 + log({\Opp}/{\Hp}). The ``UV abundance'' is derived using the dust-corrected \Oiiin~/~H$\beta$ ratio with \Teuv. The optical abundance is from the \Oiiin~/~H$\beta$ ratio using \Teopt, {equivalent to} the classical direct-$T_e$ method performed with optical spectra. For our sample, the optical and UV metallicities agree with each other to within 0.1 dex. This is a direct result of a nearly 1-to-1 ratio in the $T_{e, 1666}$ vs. $T_{e, 4363}$ relation {(Figure~\ref{fig:TePlot})}. 
We additionally report third set of 12 + log({\Opp}/{\Hp}) values considering the effect of temperature fluctuations (see \S\ref{sec:t2} for details). The results of all three methods are provided in Table~\ref{tab:abd}. 

\begin{table}
	\centering
	\caption{ Ion abundances 12+log({\Opp}/{\Hp}) measured from {\Oiiin} / {\Hb} using different $T_e$ values. In the absence of temperature fluctuations, we would expect {\Teopt} = {\Teuv} = $T_0$, with the results reflecting the true abundance for the region. Temperature fluctuations result in the abundance being higher for the $T_0$ case. Results where the $T_0$ abundance is lower are formally unphysical (corresponding to a negative variance in temperature).}
	\label{tab:abd}
	\begin{tabular}{l|ccc} 
        \hline	\hline
		& \multicolumn{3}{|c}{\bf \normalsize 12 + log({\Opp}/{\Hp})}\\
		\hline
        {\bf Object \hspace{2.5mm}} & \small {\Teopt} & \small {\Teuv} & 
         $T_0$ \\
		\hline
		SB2 & 7.68 $\pm$ 0.03 & 7.73 $\pm$ 0.02 & 7.57 $\pm$ 0.14\\
		SB82 & 7.87 $\pm$ 0.03 & 7.85 $\pm$ 0.03 & 7.95 $\pm$ 0.21\\
		SB182 & 7.93 $\pm$ 0.02 & 8.01 $\pm$ 0.05 & 7.79 $\pm$ 0.17 \\
		\hline
	\end{tabular}

\end{table}

\section{Discussion}
\label{sec:discussion}

\subsection{Temperature Fluctuations}
\label{sec:t2}

Since the UV \Oiiib~$\lambda\lambda$1661,1666 and the optical \Oiiin\ and 4363 lines have different sensitivity to $T_e$, the discrepancy between the measured $T_e$ in these lines can in principle be used to determine the temperature fluctuations.  

We follow the formalism of \citet{Peimbert67} to derive the strength of temperature fluctuations. In this framework, the temperature structure of a gaseous nebula can be characterized by two parameters: the average temperature $T_0$ and the dimensionless root mean square temperature fluctuation $t$, where $t^2$ is then the temperature variance \citep{Peimbert2013}:
\begin{subequations}
\begin{align}
    & \label{eq:T0} {T_0(\mathrm{O^{++}}) =
    \frac{\int_{}^{} T_e n_e n(\mathrm{O^{++}})dV}
    {\int_{}^{} n_e n(\mathrm{O^{++}})dV}, }\\[5pt]
    & \label{eq:t2int} {t^2(\mathrm{O^{++}}) =
    \frac{\int_{}^{} (T_e - T_0(\mathrm{O^{++}}))^2 n_e n(\mathrm{O^{++}})dV}
    {T_0(\mathrm{O^{++}})^2 \int_{}^{} n_e n(\mathrm{O^{++}})dV}. }
\end{align}
\end{subequations}

For a single ion (in this case \Opp), the $T_0$ and $t^2$ can be derived from two independent measurements of $T_e$. 
Essentially, transitions from higher energy levels (e.g., \Oiiiu\ compared to \Oiiit) are biased toward higher temperatures, resulting in higher $T_e$ derived using the standard approach (i.e., \Teuv~$>$~\Teopt\ when $t^2 > 0$). This difference in derived $T_e$ increases with $t^2$.
Assuming that the level of temperature fluctuations is relatively small compared to the average temperature $T_0$ (i.e., $t^2 \ll 1$), the measured electron temperatures can be Taylor expanded around the $T_0$, and the relationship between $t^2$, $T_0$ and $T_e$ can be expressed as
\begin{equation}
\label{eq:Tet2}
    T_{e,(\lambda1/\lambda2)} = 
    T_0\left[1 + T_0
    \left(\frac{\frac{\varepsilon^{''}_{\lambda1}(T_0)}{\varepsilon_{\lambda1}(T_0)}-\frac{\varepsilon^{''}_{\lambda2}(T_0)}{\varepsilon_{\lambda2}(T_0)}}
    {\frac{\varepsilon^{'}_{\lambda1}(T_0)}{\varepsilon_{\lambda1}(T_0)}-\frac{\varepsilon^{'}_{\lambda2}(T_0)}{\varepsilon_{\lambda2}(T_0)}}\right)
    {\frac{t^2}{2}}\right]
\end{equation}
\normalsize
\\
as derived by \cite{Peimbert2013}. 
Here $T_{e, (\lambda1~/~\lambda2)}$ represents the electron temperature measured from the flux ratio of two emission lines, $\lambda1$ and $\lambda2$. {The quantity} $\varepsilon_{\lambda 1} (T_0)$ is the emissivity of $\lambda 1$ at temperature $T_0$. $\varepsilon^{'}_i$ and $\varepsilon^{''}_i$ represent the first and second derivatives of $\varepsilon$ as a function of $T_e$.  

{
For collisionally excited lines, the emissivity as a function of $T_e$ can be approximated as
\vspace{0.5mm}
\begin{equation}\label{eq:emissivity}
    ~~\varepsilon_{\lambda}(T_e) = C_{\lambda} T_e^{-A_\lambda} \mathrm{exp}(-B_\lambda/T_e)~,
\end{equation}
where $A_{\lambda}$, $B_{\lambda}$ and $C_{\lambda}$ are coefficients for the emission line represented by $\lambda$ \citep{osterbrock06}. We find the coefficients for each line of interest by fitting Equation~\ref{eq:emissivity} to the emissivity curve derived from \textsc{PyNeb} within the range $T_e = 5000-20000$~K with $n_e = 177$ cm$^{-3}$ (i.e., the average $n_e$ found in Section~\ref{sec:Te}). The value of $n_e$ has negligible effect given the high critical densities of the relevant lines. 
For the \Opp\ lines of interest in this work, we find that $A_\lambda = 0.4$ provides an adequate approximation and adopt this as a fixed parameter. The assumed $A_\lambda$ and best-fit $B_\lambda$ and $C_\lambda$ are listed in Table~\ref{tab:emissivitycoeff}. {The fitted functions reproduce the emissivities from \texttt{PyNeb} within 2\%.}
}

\begin{table}
    \centering
    \caption{Emissivity coefficients $A_i$, $B_i$, and $C_i$ for each transition.} 
    \label{tab:emissivitycoeff}
    \begin{tabular}{l|c|c|c}
        \hline\hline
        Transition ($i$) & $A_i$ & $B_i$ (K) & $C_i$ (erg~$s^{-1}$K$^{-A_i}$~cm$^{-3}$) \\
        \hline
        \Oiiiu & 0.4 & 86,351 & 222,301\\
        \Oiiit & 0.4 & 63,417 & 22,048\\
        \Oiiin & 0.4 & 30,115 & 801.04\\
        \hline
    \end{tabular}\\
    \footnotesize{
    \begin{flushleft}    
    \textit{Notes--} Coefficients were found assuming $n_e = 177 \,\mathrm{cm^{-3}}$ (average electron density of the three objects used in this work; see Section~\ref{sec:Te}). We note that the $C_i$ obtained are listed here but are not used directly in analysis, as the $C_i$ values for each transition cancel with one another in the final relationship as expressed in Equation~\ref{eq:Tet2} and \ref{eq:t2}.
    \end{flushleft}
    }
\end{table}

\begin{figure*}
    \centering
    \includegraphics[width=.93\columnwidth]{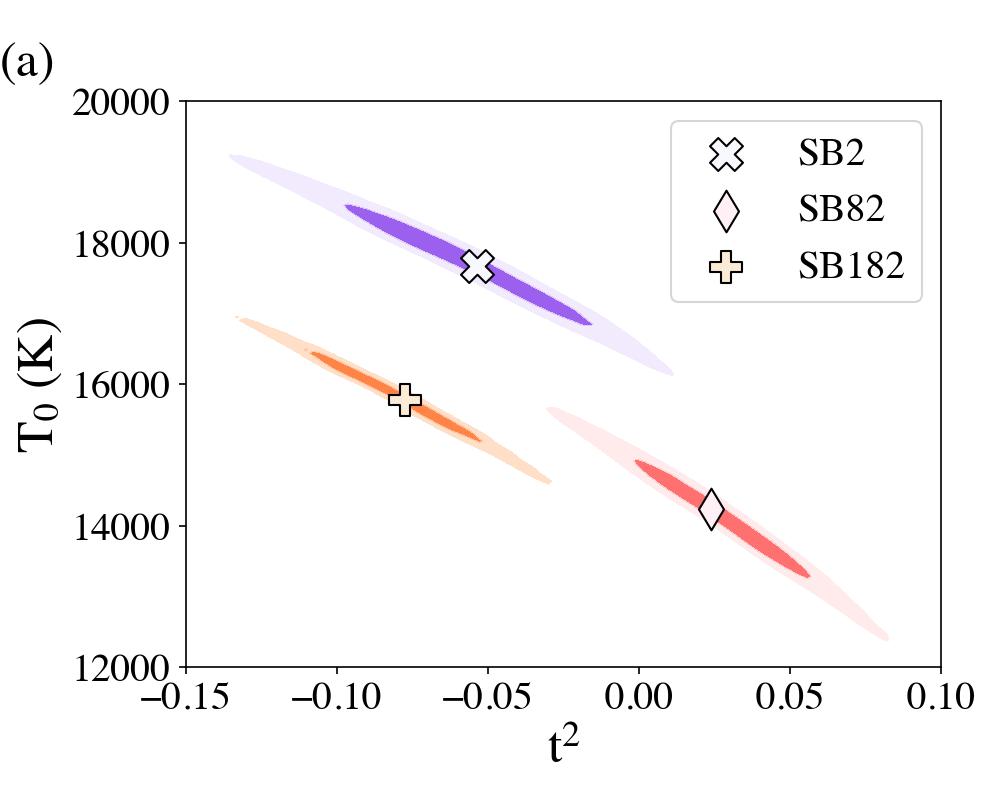}
    \hspace{5mm}
    \includegraphics[width=\columnwidth]{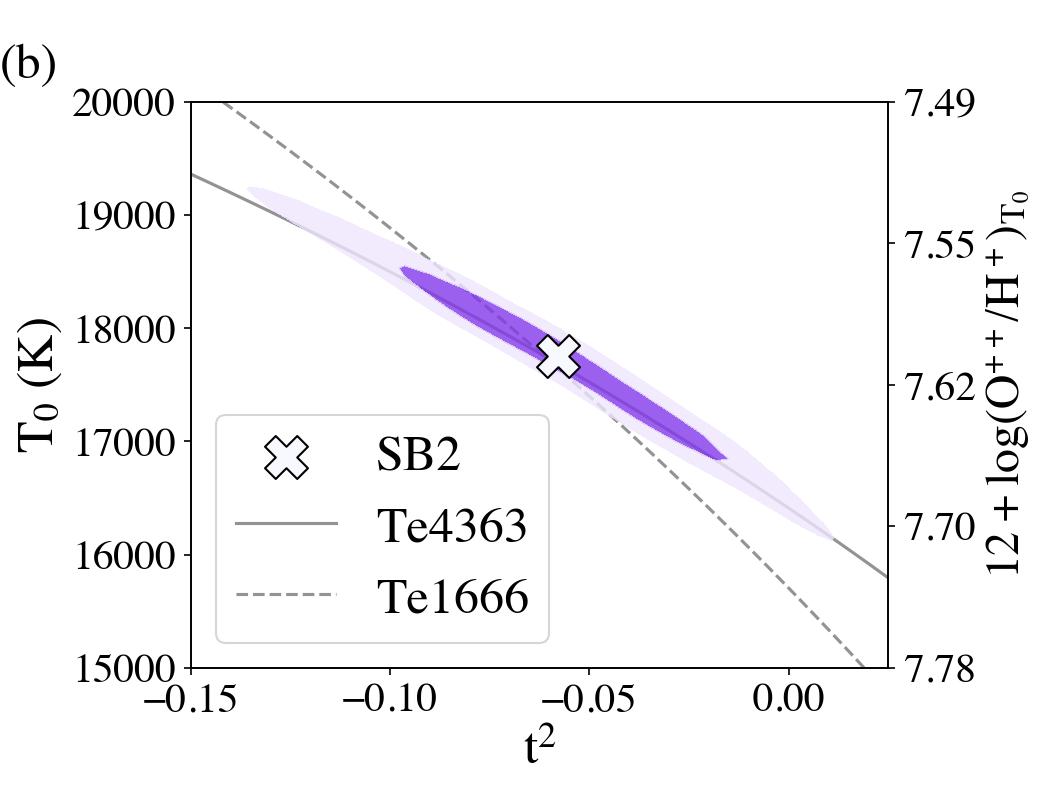}\\
    \includegraphics[width=\columnwidth]{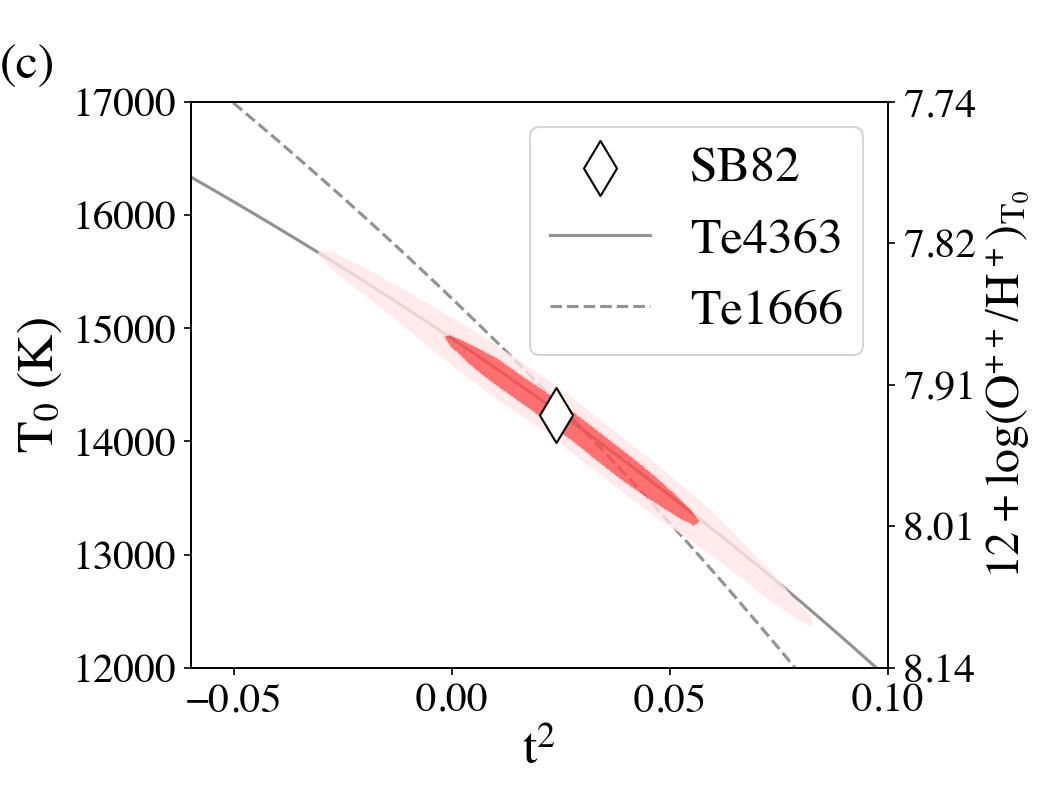}
    \includegraphics[width=\columnwidth]{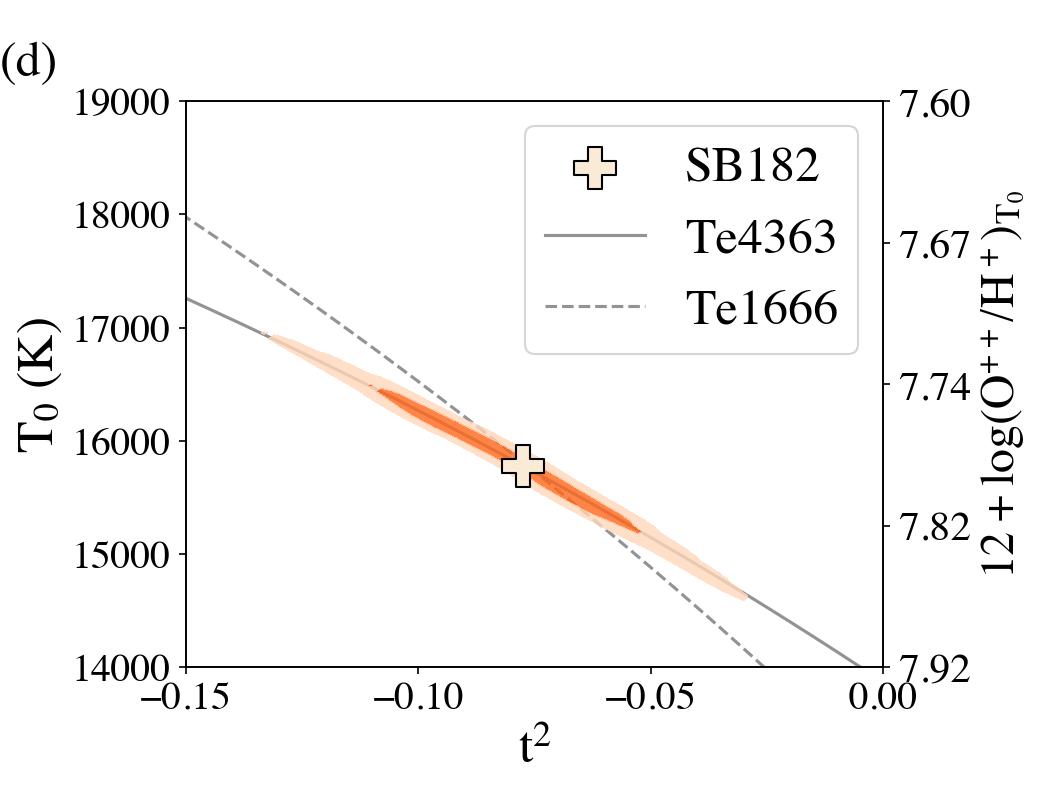}
    \caption{ The posterior of $T_0$ vs. $t^2$ from the MCMC analysis for the three BCDs in our sample. Panel (a) shows the three BCDs together, denoted by different colors and symbols. The shaded regions indicate the 1$\sigma$ and 2$\sigma$ confidence intervals. 
    Panels (b)--(d) zoom in on the individual objects. For each object, the secondary y-axis shows the $\mathrm{12+log_{10}}$({\Opp}/{\Hp}) abundance at the corresponding $T_0$. The solid and dashed lines represent the relation between $T_0$ and $t^2$ from a single $T_{e, 4363}$ or $T_{e, 1666}$. 
    We note that the relationship between $T_e$ and $t^2$ (Eq. \ref{eq:t2}), as derived from Eq. \ref{eq:Tet2}, is the mathematical reason for the negative value for $t^2$. However, it is not physically possible for {\Teuv} to be less than {\Teopt} in the presence of temperature fluctuations. }
    \label{fig:t2T0_ind}
\end{figure*}

{With the coefficients in Table~\ref{tab:emissivitycoeff}, we apply Equations~\ref{eq:Tet2} and \ref{eq:emissivity} with the measured values of \Teuv\ and \Teopt\ from our targets. 
The dimensionless variance $t^2$ is directly proportional to the discrepancy between these two temperatures, and can be expressed as}
\begin{equation} \label{eq:t2}
    ~~t^2~(T_{e~4363},T_{e~1666})~ = \frac{2}{B_{4363}-B_{1666}}\left[T_{e~4363}-T_{e~1666}\right].
\end{equation}
The {derived values} of $t^2$ and $T_0$ are presented in Table~\ref{tab:Te}. 
Due to the correlation between derived $T_0$ and $t^2$, we use the Markov Chain Monte Carlo (MCMC) method to estimate the uncertainties associated with $t^2$ and $T_0$. {We simultaneously account for} systematic uncertainties arising from flux calibration and {attenuation} correction described earlier.
Our MCMC analysis is constructed to fit the observed \Oiiin, \Oiiit, and \Oiiiu\ fluxes. The free parameters are $T_0$, $t^2$, and a flux normalization factor (e.g., relative to \Heii\ emission). The E(B-V) values and errors derived from \Hi\ Balmer and Paschen lines (see Section~\ref{sec:OptReddening}) are included as priors to model the optical \Oiiin\ and \Oiiit\ fluxes. For UV \Oiiiu\ emission, the error budget is dominated by the reliability of using \Heii\ flux ratios to correct the aperture and reddening effects. This uncertainty is estimated by adding a {$\pm 4,000$~K} uncertainty in $T_e$(\Heii) when predicting the intrinsic \Heiiu~/~\Heiio\ ratio to derive the reddening- and aperture-corrected \Oiiiu\ flux (see Section~\ref{sec:UVreddening} for details). The posterior $1\sigma$ and $2\sigma$ confidence intervals for $(t^2, T_0)$ for SB~2, SB~82, and SB~182 are presented in Figure \ref{fig:t2T0_ind}. The $1\sigma$ uncertainties in $t^2$ and $T_0$ are given in Table \ref{tab:Te}.

\begin{figure*}
    \centering
    \includegraphics[width=1.7\columnwidth]{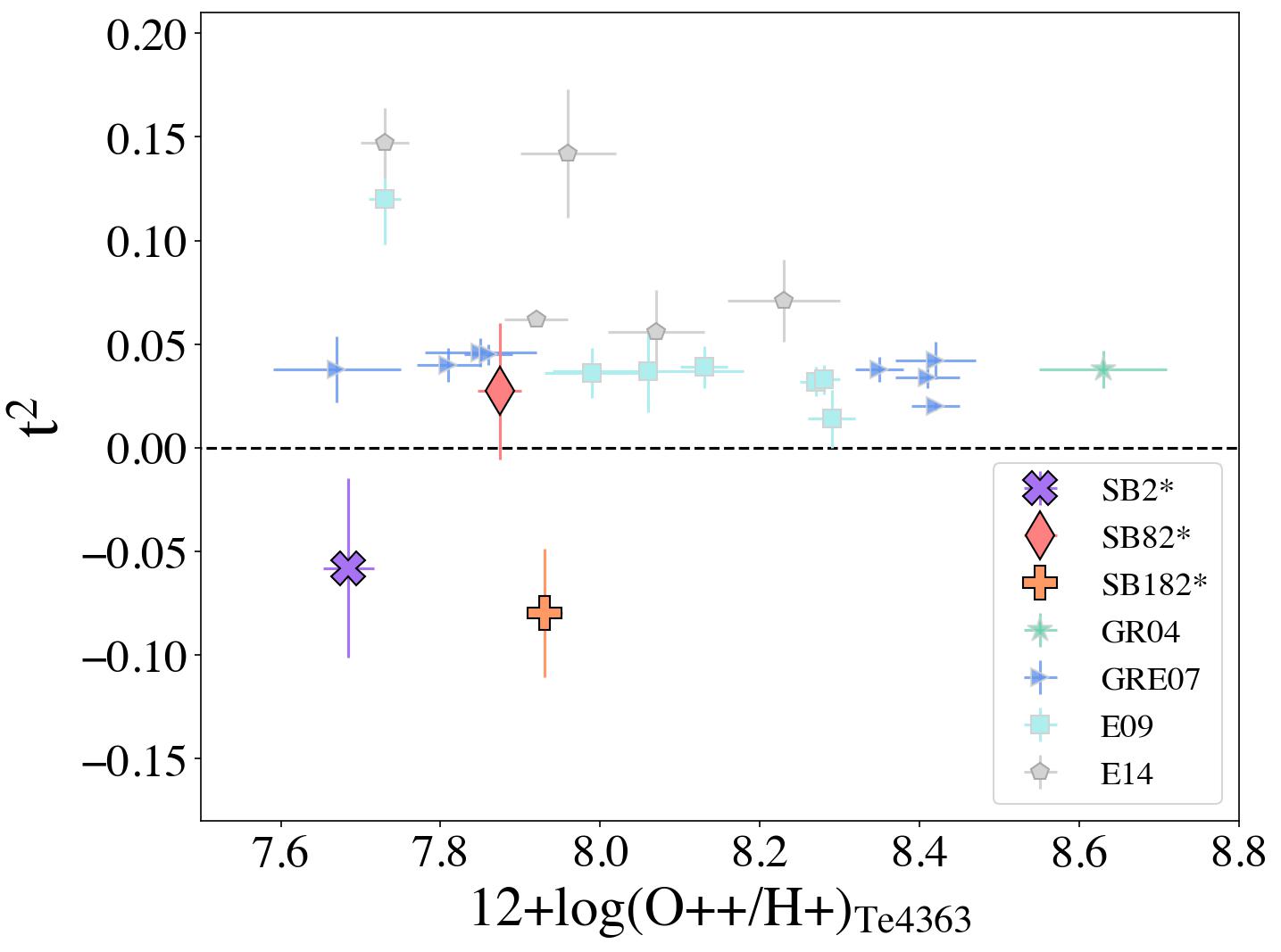}
    \caption{Temperature variance ($t^2$) and the corresponding {\Opp}/H abundances based on {\Teopt} for objects SB~2, SB~82, and SB~182 ($* =$ this work), compared to $t^2$ and {\Opp}/H abundances from the literature \citep{GarciaRojas2004,Garcia2007,Esteban2009,Esteban2014}. SB~82 displays a positive $t^2$ value within 1$\sigma$ of zero, while our $t^2$ results for SB~2 and SB~182 are negative. Formally, a negative value for $t^2$ is unphysical, as this would indicate that UV emission is less affected than optical emission at higher temperatures. However, in the temperature fluctuation paradigm, a negative $t^2$ is mathematically possible when the measured value for {\Teuv} is lower than that of {\Teopt}.
    The $t^2$ values measured directly for our sample are negative on average, and in tension with those inferred to explain the abundance discrepancy factor in the literature sample.
    }
    \label{fig:OH_t2}
\end{figure*}

{Physically we expect {$t^2 \geq\ 0$} due to temperature fluctuations, which result in measured \Teuv~$>$~\Teopt. Surprisingly, however, we find the opposite result for SB~2 and SB~182 with derived $t^2 < 0$. Only SB~82 has a best-fit solution which is physically allowed, {but is 1$\sigma$ consistent with $t^2 = 0$.} Our analysis formalism intentionally enables us to derive negative (i.e., unphysical) $t^2$ values, in order to assess uncertainties and obtain accurate sample averages.
As a point of comparison, in Figure~\ref{fig:OH_t2} we plot $t^2$ {derived directly from} our sample's {$T_e$ measurements} with that inferred for a broader literature sample based on the abundance discrepancy factor (i.e., the $t^2$ needed to reconcile abundances measured from optical CELs and recombination lines). If temperature fluctuations are the dominant cause of the abundance discrepancy, we expect these two methods of obtaining $t^2$ to agree within the uncertainties. 
This is clearly not the case for SB~2 and SB~182, which differ by several standard deviations from the typical values $t^2 \approx 0.02$--0.1 inferred from ADF measurements, although SB~82 is in reasonable agreement. 
Our sample of three targets has an unweighted mean $\left< t^2 \right> = -0.04 \pm 0.03$ (with sample standard deviation 0.05, larger than the measurement uncertainties). 
Notably the precision in $t^2$ measured with our method is comparable to that from ADF-based results. From the results shown in Figure~\ref{fig:OH_t2}, we conclude that our direct measurements of $t^2$ are generally inconsistent with the values needed to explain the ADF observed in \Hii\ regions of similar metallicity.}

While our results are in tension with the standard explanation of $t^2$ as the cause of the ADF, they are consistent with the trend of UV- and optical-based temperatures found by \citetalias{Mingozzi_etal_2022}.
Specifically \citetalias{Mingozzi_etal_2022} measured both \Teopt\ and \Teuv, and found that they lie approximately on a 1:1 trend.
Our measurements fall well within the scatter of the \citetalias{Mingozzi_etal_2022} results, as shown in Figure~\ref{fig:TePlot}. Two of the objects studied in this work, SB~2 and SB~182, were also analysed in \citetalias{Mingozzi_etal_2022} and are those for which we report negative values of $t^2$. For these two objects, our electron temperature measurements are consistent within $1\sigma$ of those reported in \citetalias{Mingozzi_etal_2022}.
When accounting for all three objects in this work, our results show a tighter correlation to the 1:1 line than \citetalias{Mingozzi_etal_2022}'s distribution, potentially due to the difference in how the {UV dust attenuation is accounted for}.
The lower scatter using the \Heii\ method is promising, and we suggest further analysis involving a larger sample of BCDs with optical and UV \Heii\ emission to explore the potential of the \Heii\ method as a calibration tool in optical+UV emission line analyses.

As they stand, our $t^2$ results do not support the temperature fluctuation framework as the physical driver behind the ADF, and an alternate explanation is therefore needed.
A key result is that SB~2 and SB~182 exhibit \Teuv\ measurements which are \textit{lower} than \Teopt\ -- an unphysical phenomenon. Effects such as those described by the $\kappa$ distribution \citep[e.g., ][]{Nicholls2012} or the multi-component thermal electron distribution model \citep{Storey2014} can be ruled out, since both would enhance the population of all high-energy levels rather than exclusively the 5.35 eV level.
An alternate explanation for the ADF is the metal-rich inclusion hypothesis \citep[e.g.,][]{Liu2000,Tsamis2003,Stasinska2007}. 
{In this scenario, small-scale metal-rich droplets are interspersed throughout the {\Hii} region. However, since the droplets are proposed to be cooler and would have essentially no contribution to optical and UV CEL emission lines, including the \Oiiit\ and \Oiiiu\ emission that are being tested in this work, our test does not provide sufficient constraints on this scenario.  }

{In the next section, we further examine the reliability of our analysis and discuss the implications of the unphysical $t^2$ results for SB~2 and SB~182.}

\subsection{Examination of Unphysical $t^2$ Results}
\label{unphysical}

We now consider possible effects which could cause the unphysical $t^2 < 0$ values measured for two of the three targets in our sample. \Hii\ regions have inherently complex structure, and various physical effects can affect the measurements presented in this work. 
Temperature fluctuations should lead to derived \Teuv~$\geq$~\Teopt\ (corresponding to $t^2 \geq 0$; Equation~\ref{eq:t2}).
Our finding of $t^2 < 0$ suggests that some other process may be reducing the observed flux ratio of \Oiiiu/\Heii~1640 used to derive \Teuv\ (or alternatively, increasing the optical \Oiii/\Heii~4686 ratios).
In particular, the flux ratio \Oiiiu/\Heii~1640 would need to increase by $\simeq$10\% for SB~2 and SB~182 in order to obtain a result of $t^2 = 0$. Here we examine the extent to which different spatial distributions of \Oiii\ and \Heii\ emission regions, combined with differential dust attenuation or aperture effects, can impact the results.

We first consider the effects of differential dust obscuration. UV emission is especially sensitive to dust attenuation.
\citet{Chen2024} found that $\sim 50\%$ of the optical \Oiii\ emission could be missing compared to the far-IR \Oiii\ fine-structure emission. A similar mechanism may preferentially occlude \Oiiiu\ relative to \Heii.
Since \Opp\ and \Hepp\ are not strictly co-spatial {due to having different ionization energies (35~eV vs. 54~eV)}, {it is possible that the \Oiiiu\ and \Heiiu\ lines have different attenuation as well}. We assess the degree of differential attenuation by comparing optical diagnostics arising from different ionization zones. In addition to the \Hi\ Balmer and Paschen lines reported in Table~\ref{tab:flux}, we also detect \Hei\ recombination lines in all targets as well as the optical \Heii\ line \Heii~5411 in SB~2.
For SB~2, we derive E(B-V)~$=0.148\pm0.021$ from \Hei~6678/\Hei~5876, which is considered one of the most robust \Hei\ line ratios (\citealt{mendez-delgado-hei}), and $0.25\pm0.11$ from \Heii~5411/\Heii~4686, compared with $0.167\pm0.017$ from \Hi. These are consistent within 1$\sigma$, with no evidence for less attenuation of \Heii, though we note the increased uncertainty for E(B-V) measured using \Heii. For SB~182 we derive E(B-V)~$0.009\pm0.015$ from \Hei~5876/\Hei~6678, in agreement with the value $0.001\pm0.012$ from \Hi. 
We find similar consistency for SB~82, with E(B-V)~$0.042\pm0.018$ from \Hei\ and $0.035\pm0.016$ from \Hi, although we note that this object does not have unphysical results. 
We can also estimate the reddening in the \Oiii-emitting zone by using the nearly uniform intrinsic flux ratio of \Neiii~3869/\Oiii~5007~$=0.082$, with intrinsic scatter of only 0.04 dex \citep{Jones2015}. In all cases the observed \Neiii/\Oiii\ is consistent with that expected from the \Hi-based reddening. Moreover, multiple diagnostics indicate that the reddening is relatively low in all targets. 
We conclude that there is no evidence for differential dust attenuation which could cause increased reddening of the \Oiii\ relative to \Heii\ emission, and that differential dust attenuation is not responsible for the unphysical $t^2$ measurements in SB~2 or SB~182.

We now consider whether different spectroscopic apertures, in combination with spatial structure of the targets, could cause the derived negative $t^2$ values in SB~2 and SB~182.
This would require a scenario where the physical region probed by COS in the UV is characterized by an \Oiiiu/\Heii\ flux ratio which is $\sim$10\% lower than the region covered by ESI optical spectroscopy (specifically, $<$10\% lower for SB~2 and 13\% lower for SB~182).
We note that \Heii\ emission morphology is often more compact and closer to the ionizing source compared to typical nebular emission from \Oiii\ \citep[e.g.,][]{kehrig18, crystalmartin-peng-li2024, rickards-vaugt21}. 
For COS spectra, spatial extent in the dispersion direction can result in broader emission line profiles. We thus checked the spectral FWHM of \Heiiu\ and \Oiiiu, and found  that \Heiiu\ is marginally broader than \Oiiiu\ in all three objects ($8\%$ broader in SB~2 and SB~182, and $37\%$ broader in SB~82), which may arise from intrinsic kinematic differences or alternatively a larger spatial extent of \Heiiu\ {that may be possible with strong Wolf-Rayet stellar winds}.
To more directly quantify the magnitude of aperture effects, we extract the COS spectra using different ``BOXCAR'' extraction heights. We consider a 10-pixel height which is similar to the $1.0~''$ wide ESI slit, and a 35-pixel height which captures all light in the $2.5~''$ COS aperture.
The 35 pixel height is expected to be similar to the TWOZONE extraction used for our primary results, and indeed in all cases they are consistent within the 1$\sigma$ measurement uncertainties.
For SB~2, we find that a 10 pixel BOXCAR extraction height results in an 11\% larger \Oiiiu/\Heii~1640 ratio compared to the TWOZONE extraction, yielding $t^2 = -0.02\pm0.02$. For SB~82 and SB~182, all extraction heights yield \Oiiiu/\Heii~1640 ratios which are consistent within the 1$\sigma$ uncertainties ($<$5\% and $<$3\% difference, respectively). 
We find that SB~2 is consistent at the 1-$\sigma$ level with the case of uniform temperature ($t^2 = 0$) when considering possible aperture effects, while the negative $t^2$ result for SB~182 is not explained by aperture effects. 
We caution that this analysis of COS aperture effects is subject to complications arising from vignetting and other optical properties \citep[e.g.,][]{James2022}. However, the targets in this work are relatively compact such that we expect vignetting and light leakage beyond the nominal 1.25\arcsec\ COS aperture to be minimal. Ultimately, future integral field spectroscopic observations at both UV and optical wavelengths will be ideal to provide robust aperture matching to resolve this possible issue.

{In summary, we investigated whether dust attenuation and aperture effects can plausibly explain the unphysical temperature structure measured for our targets SB~2 and SB~182 (i.e., \Teuv~$<$~\Teopt, and $t^2 < 0$ in our formalism). For SB~2, correcting for aperture effects can bring our measurement within $1\sigma$ of $t^2 = 0$, although we still formally find a negative $t^2$. 
We find no plausible explanation for SB~182, which shows no significant spatial variations and has reddening consistent with zero. We likewise find no significant change for SB~82 where we derive a slightly positive $t^2$. 
Considering these possible systematic uncertainties, we still find an average $t^2 \lesssim 0$ for our aggregate sample which suggests little or no temperature fluctuations.}

\subsection{Implications and Future Prospects}

Our results {show} generally good agreement ($<$ 0.1 dex) between the $T_e$ and metallicity measured from the optical and UV CELs for \Opp.
This is promising for obsesrvations at cosmic dawn ($z\gtrsim 6$), as recent observations of such high-z galaxies from JWST rely on rest-UV nebular emission features to probe their physical properties \citep[e.g., ][]{rhoads23, ArellanoCordova2022, Jones2023, wangx24, cameron23, topping24, topping22, leethochawalit22}. 
At least for local blue compact dwarf galaxies, our results show that the \Oiii\ emission lines between the optical and UV are consistent with each other within 550~K ($\sim$3.5\%) on average.
Additionally, our results are consistent with the findings of \cite{bresolin2016}, who found that for low-metallicity systems, $T_e$-based abundances are in relatively good agreement with the stellar metallicities {of young B supergiants} accounting for a $\sim 0.1$ dex correction to nebular oxygen abundance due to dust depletion.

With continuous spectral coverage from the rest UV to rest optical of galaxies at $z\sim5-9$, JWST/NIRSpec also offers a unique opportunity to reduce the aperture matching uncertainties and to investigate the reliability of using \Heii\ emitting galaxies for similar analyses in the future. Our work using \Heii\ to calibrate optical and UV spectra shows that the \Heii\ method has potential for increasing the precision of \Hii\ region gas-phase diagnostics.
We additionally note that in measurements of $t^2$, the posteriors in the distribution follow \Teopt\ (see Figure \ref{fig:t2T0_ind}), with the error budget in $t^2$ being dominated by the error on {\Teuv}. The error on \Teuv\ is largely systematic (arising from the \Heii\ correction of the UV oxygen lines), which indicates a promising prospect for future observations as future targets may not need such high SNR as the objects presented in this work.
More comparisons, especially at various redshifts, would collectively yield stronger constraints on $t^2$ and the systematic uncertainties. 

In addition to the traditional 1D spectral analysis, recent observations using integral field units (IFU) have proven to be valuable in determining the spatial structure of ionized nebulae, especially the spatial extent of different ions \citep[e.g.,][]{kehrig18, mcleod19, kumari17, Westmoquette07, fensch16, lago19, Weilbacher15, cosens22}. Similar observations focusing on \Heii\ emitting galaxies would help us determine the systematic bias of our method.
We additionally note that the three objects studied in this work do not have a measured ADF as we do not have measurements of their oxygen RLs.
Knowing the magnitude of the ADF is not strictly necessary for determining the magnitude of temperature fluctuations or making direct $T_e$ measurements, but would certainly aid with interpreting the results. Having a known ADF would be valuable context for the measured values of $t^2$, and to test whether our direct measurements of $t^2$ correlate with those inferred from the ADF. 
As such, follow up to obtain RLs and to measure the ADF (or lack thereof) in SB~2, SB~82, and SB~182 would be a natural next step in order to better understand the physical driver behind the ADF.

\section{Summary}
\label{conclusion}

In this work, we present a novel method that uses the nebular \Heiiu\ and \Heiio\ emission lines to perform aperture and reddening correction between UV (HST/COS) and optical (Keck/ESI) spectra, facilitating precise direct comparisons for \Opp/\Hp\ abundances and $T_e$ measured between the optical and UV for three nearby BCDs. Our main findings are:
\begin{itemize}
    \item UV and optical $T_e$ are consistent with each other within $\sim 500$~K (standard deviation of scatter). This result is consistent with the similar analysis by \citetalias{Mingozzi_etal_2022} using SED models for aperture matching, but appears to be closer to the 1-to-1 ratio and exhibits smaller scatter, indicating that our method could improve the precision in matching the UV- and optical-based nebular properties for similar galaxies.
    \item Similarly, the UV and optical \Opp/\Hp\ metallicities are consistent with each other within 0.1 dex, verifying that for local dwarf galaxies, the gas-phase metallicities measured between the UV and optical CEL methods can be used interchangeably, and that UV CELs can be reliably used to trace oxygen abundances using the $T_e$ method.
    \item Two galaxies (SB~2 and SB~182) have \Teuv$<$\Teopt, and UV-based 12+log(\Opp/\Hp) which is larger than the optical-based value, an unphysical result which contradicts predictions from temperature fluctuation scenarios which are often invoked to explain the abundance discrepancy factor.
    The unphysical $t^2$ value for SB~2 can be attributed in part to possible aperture effects. For SB~182, the physical cause behind the unphysical $t^2$ value remains an open question.
\end{itemize}

This work successfully validates the feasibility of using nebular \Heii\ emission to match the UV and optical spectra observed from different instruments. However, with only three galaxies in the sample, this work is subject to small-sample statistical uncertainties and sample variance. Similar observations of a larger sample of \Heii\ emitting galaxies, as well as acquiring ADF measurements for the three objects studied in this work, would significantly improve our understanding of potential systematic uncertainties and help to address the physical cause for objects with derived $t^2<0$. Additionally, observations from JWST and optical IFSs on \Heii\ emitting galaxies can help to further validate and understand the systematic biases which may be inherent in this methodology.
Overall, our results establish a clear foundation for using rest-frame optical and UV emission line diagnostics interchangeably, enabling robust studies of chemical evolution and other galaxy properties based on these features, applicable from the present-day universe to the most distant objects being studied with JWST.

\section*{Acknowledgements}

This research is based largely on observations with the NASA/ESA Hubble Space Telescope obtained from the Space Telescope Science Institute, which is operated by the Association of Universities for Research in Astronomy, Incorporated, under NASA contract NAS5-26555. 
These observations are associated with programs HST-GO-15646 and HST-GO-14168. 
Support for program number HST-GO-15646 was provided through a grant from the STScI under NASA contract NAS5-26555.
TJ acknowledges support from the NASA under grants HST-GO-15646, HST-GO-16697, and 80NSSC23K1132, and from a UC Davis Chancellor's Fellowship.
YC is supported by the Direct Grant for Research (C0010-4053720) from the Faculty of Science, the Chinese University of Hong Kong.
The optical spectroscopic data presented herein were obtained at the W. M. Keck Observatory, which is operated as a scientific partnership among the California Institute of Technology, the University of California and the National Aeronautics and Space Administration. The Observatory was made possible by the generous financial support of the W. M. Keck Foundation.
The authors wish to recognize and acknowledge the very significant cultural role and reverence that the summit of Mauna Kea has always had within the indigenous Hawaiian community.
We are deeply appreciative of the opportunity to conduct observations from this mountain. 
In addition, EH would like to give special thanks to Kelsey Glazer and Dr. Erika Holmbeck, who made the completion of this work possible.

\section*{Data Availability}

The HST/COS spectra used in this article are publicly available at Mikulski Archive for Space Telescopes (MAST) Portal. The raw Keck/ESI observations are available at the Keck Observatory Archive (KOA). The reduced spectra can be shared by the authors upon reasonable request.

\bibliographystyle{apj}
\bibliography{bib.bib} 

\end{document}